\documentclass[useAMS,usenatbib]{mn2e}

\usepackage{epsfig}
\renewcommand{\vec}[1]{\mbox{\boldmath $#1$}}

\title[Polarization of the Crab Nebula]
{Polarization of the Crab Nebula with disordered magnetic components}

\author[Y.~Nakamura, S.~Shibata]
  {Y.~Nakamura,$^1$\thanks{yuji@ksirius.kj.yamagata-u.ac.jp}
  S.~Shibata,$^2$\thanks{shibata@sci.kj.yamagata-u.ac.jp} \\
  $^1$Graduate School of Science and Engineering, Yamagata University, Yamagata
990-8560, Japan\\
  $^2$Department of Physics, Yamagata University, Yamagata 990-8560, Japan }

\date{Accepted 2007 June 28. Received 2007 June 21; in original form 2007 February 6}

\pagerange{\pageref{firstpage}--\pageref{lastpage}} \pubyear{2007}

\def\LaTeX{L\kern-.36em\raise.3ex\hbox{a}\kern-.15em
    T\kern-.1667em\lower.7ex\hbox{E}\kern-.125emX}

\begin{document}
\label{firstpage}

\maketitle

\begin{abstract}
In this paper, we present an expanding disc model to derive polarization properties of the Crab nebula.
The distribution function of the plasma and the energy density of the magnetic field are prescribed 
as function of the distance from the pulsar 
by using the model by Kennel and Coroniti (1984) with $\sigma = 0.003$, 
where $\sigma$ is the ratio of Poynting flux to the kinetic energy flux 
in the bulk motion just before the termination shock. 
Unlike previous models, we introduce disordered magnetic field, 
which is parameterized by the fractional energy density of the disordered component. 
Flow dynamics is not solved. The mean field is toroidal.

Averaged polarization degree over the disc is obtained as a function 
of inclination angle and fractional energy density of the disordered magnetic field.
It is found for the Crab that the disordered component has about $60$ percent of the magnetic field energy. 
This value is also supported by the facts 
that the disc appears not `lip-shape' but as `rings' in the intensity map as was observed, 
and that the highest polarization degree of $\sim 40$ percent is reproduced for rings, 
being consistent with the observation.

We suggest that because the disordered field contributes rather pressure than tension, 
the pinch force may be over-estimated in previous relativistic magnetohydrodynamic simulations. 
Disruption of the toroidal magnetic field with alternating direction, 
which is proposed by Lyubarski (2003), may actually takes place. 
The relativistic flow speed, 
which is indicated by the front-back contrast, 
can be detected in asymmetry in distributions of the position angle and depolarization.
\end{abstract}


\section{Introduction}
\label{intro}
It is well established that the Crab Nebula shines in synchrotron radiation. 
However the origin of the high energy particles has been a long-standing problem.
The pulsar wind from the central pulsar is certainly the source of the particles and the magnetic fields.
We nevertheless do not know what fraction of the wind energy
is in magnetic field and in plasma kinetic energy. 
We sometimes introduce the $\sigma$-parameter, which is the ratio of
the magnetic energy flux to the kinetic one of the wind bulk motion just before the termination shock.

Kennel and Coroniti (KC) (1984a, 1984b) proposed a simple model in which
the nebula is expanding hot plasmas after the termination shock
of the wind. In their model, the kinetic energy of the wind bulk motion
converts into heat with non-thermal components.
This picture results in a small value of $\sigma$, i.e., dominance of the kinetic energy, to explain the observed
synchrotron luminosity and the presumed expansion velocity of $\sim 2000 \ {\rm km} \; {\rm s}^{-1}$.
One finds it difficult to reproduce the kinetic energy dominated wind by means of 
the axisymmetric ideal-MHD model (Michel 1969, Bogovalov 1998, Begelman \& Li 1994, Tomimatsu 1994), 
although some possibilities for the small $\sigma$ 
have been pointed out (Okamoto 2002, Vlahakis 2004). 
This is referred to as the $\sigma$-problem.

Because the magnetic axis inclines to the rotation axis,
the magnetic
neutral sheet in the equator is folded for each rotation to produce
a series of current sheets flowing out with the wind.
Some authors attempt to solve the $\sigma$-problem by introducing
dissipation of the magnetic field in such current sheets
(Coroniti 1990, Kirk \& Skjaeraasen 2003).

Lyubarski (2003) has proposed that conversion of the magnetic energy to
heat takes place when the current sheets come into the termination shock.
With this picture, Poynting energy directly changes to plasma heat and
non-thermal particles, so that
the pulsar wind should not always be kinetic energy dominant, i.e.,
the $\sigma$-problem may not be a problem.

Magnetic field of the wind is essentially toroidal far beyond the light cylinder owing to rotation.
The nebula field convected from the wind is also believed to be toroidal,
indicated by highly axisymmetric structures in the Crab Nebula.
However, if disruption of the numerous current sheets takes place at or after the termination shock, 
one may expect that the nebula field is not pure toroidal 
but it has strong disordered components.
Thus, the disordered magnetic field may be a good indicator for energy conversion 
from the toroidal magnetic field to plasmas.
To diagnose disruptive processes of the magnetic field, 
it must be very much helpful to measure magnitude of the disordered field.

Disordered magnetic field can be detected by image analysis and by polarization analysis. 
Shibata et al. (2003; Paper I) calculated images by use of the KC model and compared with the Chandra observations.
If the nebula field is pure toroidal, intensity along the major axis of the torus should be reduced, 
and the images look like `lips' rather than rings. 
The X-ray images, which appear as rings, indicate 
that disordered field may be comparable to the mean field.

Polarization observation of the nebula provides the mean field structure and degree of randomness of the field. 
The polarization of the Crab Nebula is observed in radio bands (Wilson 1972, Velusamy 1985), 
in optical bands (Oort \& Walraven 1956, Schdmidt et al. 1979, Hickson \& van der Bergh 1990, and Michel et al. 1991), 
and in X-ray bands (Weisskopf et al. 1978).
The X-ray polarization is not spatially-resolved but a mean polarization is obtained to be $\sim 20 \%$. 
Almost the same value is obtained also in optical bands (Oort \& Walraven 1956). 

In this paper, we calculate polarization properties of a nebula which expands with relativistic speeds 
and with disordered magnetic field as well as mean toroidal fields. 
We then compare the results with the Crab observations, 
so that the magnitude of the disordered magnetic field is estimated.

\section{Polarization model}
\subsection{Polarization from a medium in relativistic motion}
\label{uniB}
Let us consider synchrotron radiation from a relativistic plasma moving relative to the observer at the velocity \vec{V}.
The volume emissivities in terms of the Stokes parameters 
in the `flow frame' comoving with the plasma may be given by (Rybicki \& Lightman 1979)  
\begin{eqnarray}
\left(
\begin{array}{c}
d I^\prime / d s \\
d Q^\prime / d s \\
d U^\prime / d s \\
d V^\prime / d s
\end{array}
\right)
&=&
\left(
\begin{array}{c}
j^\prime_{tot} \\
j^\prime_{pol} \\
0 \\
0
\end{array}
\right) \nonumber \\
&=&
\left(
\begin{array}{c}
\frac{p+7/3}{p+1} ~ \Phi(\omega^\prime,p) ~ (B^\prime \sin{\theta^\prime})^{\frac{p+1}{2}}\\
\Phi(\omega^\prime,p) ~ (B^\prime \sin{\theta^\prime})^{\frac{p+1}{2}}\\
0\\
0
\end{array}
\right), \label{Stok_0}
\end{eqnarray}
where
\begin{eqnarray}
\Phi(\omega^\prime,p) = \frac{\sqrt{3} e^3 K}{8 \pi m c^2} \left( \frac{3 e}{m c \omega^\prime} \right)^{\frac{p-1}{2}} \Gamma \left( \frac{p}{4} + \frac{7}{12} \right) \Gamma \left( \frac{p}{4} - \frac{1}{12} \right), 
\end{eqnarray}
$B^\prime=|\vec{B}^\prime|$ is the magnetic field strength, 
$\Gamma$ is the gamma function, 
$-e, m$ are the charge and the rest mass of an electron, and $c$ is the speed of light.
Here, we have assumed for a given volume element in which 
the magnetic field $\vec{B}^\prime$ is uniform, 
and the relativistic electrons and positrons have a power-law energy distribution, 
$f(\gamma) = K \gamma^{-p}$, where $\gamma$ is the Lorentz factor of the particles; 
$p$ and $K$ are functions of position.
The pitch angle distribution is assumed to be uniform.
In the flow frame, the magnetic field $\vec{B}^\prime$ and the unit vector toward the observer $\vec{n}^\prime$ make the angle $\theta^\prime$, 
and the angular frequency of the radiation observed at $\omega$ is Doppler-shifted to $\omega^\prime$.

\begin{figure}
\begin{center}
\includegraphics[clip,width=8.4cm]{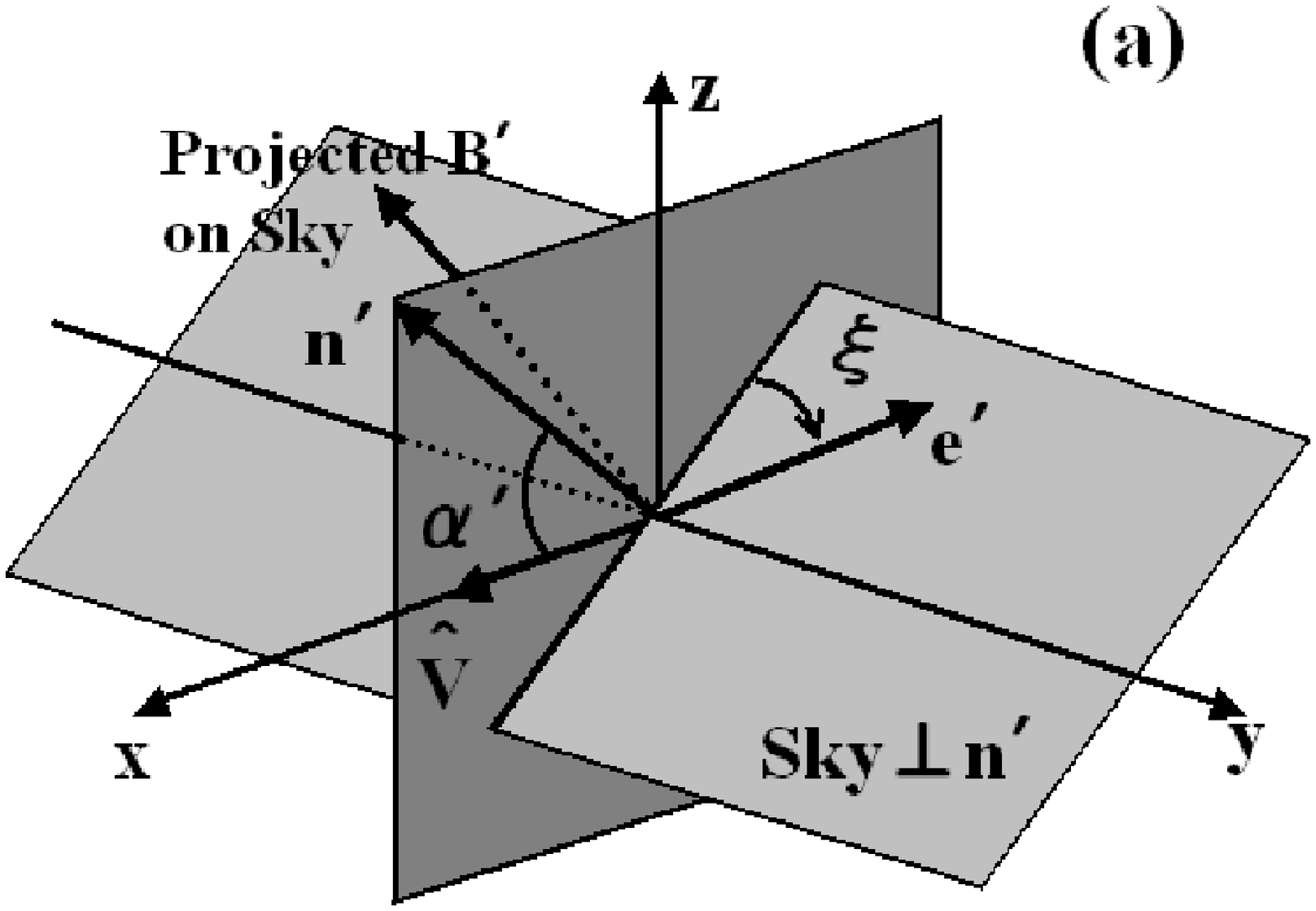}
\includegraphics[clip,width=8.4cm]{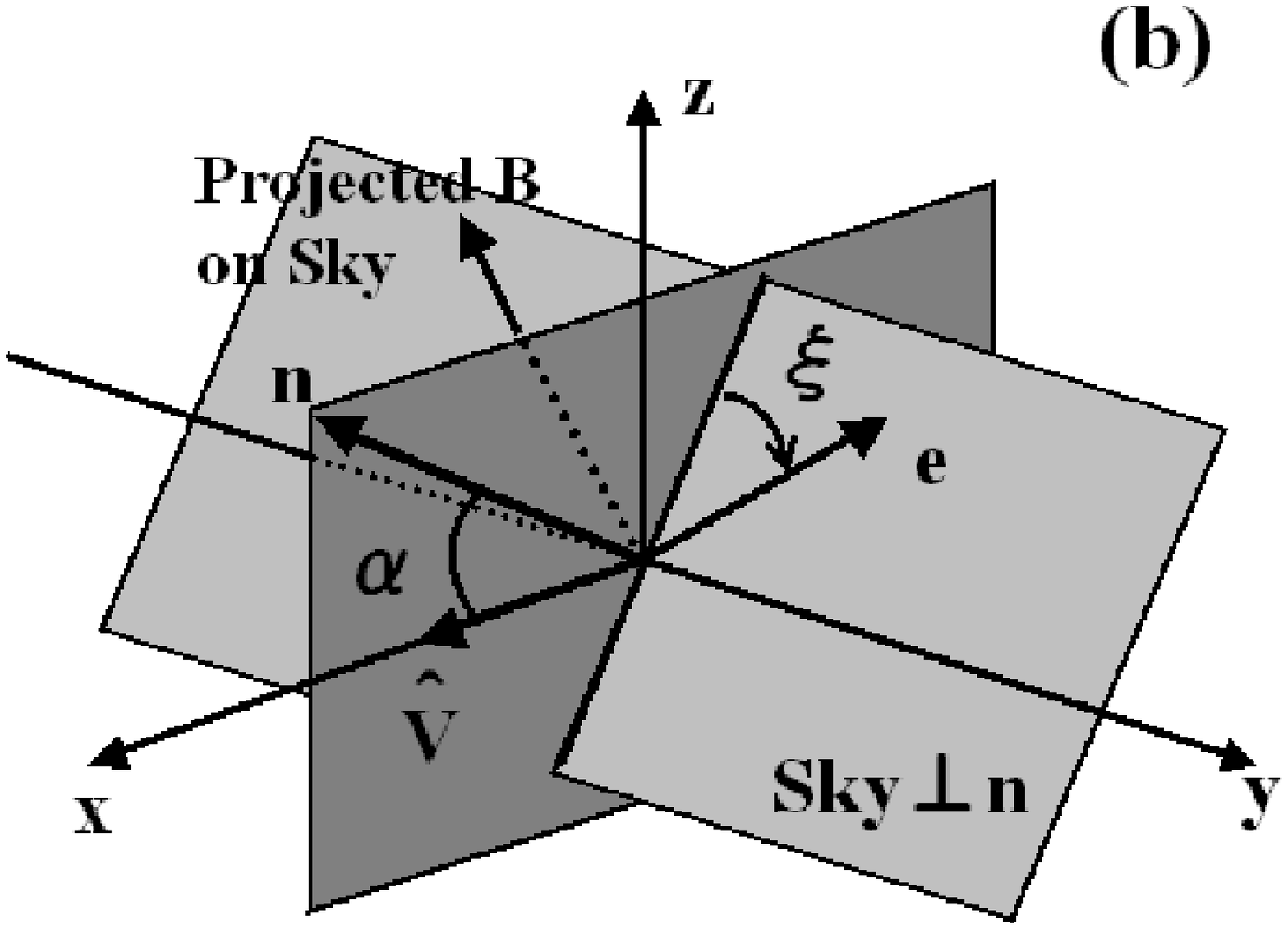}
\end{center}
\caption{ \label{flow_obs}
(a) and (b) show the flow frame and the observer frame, respectively.
The position angle $\xi$ is an invariant.
The unit vectors of the flow velocity, the observer's direction and the electric vector of the linearly polarized wave are indicated by $\hat{\vec{V}}$, $\vec{n}$ and $\vec{e}$, respectively.}
\end{figure}

If the ideal-MHD condition \vec{E} $+$ \vec{\beta} $\times$ \vec{B} = \vec{0} holds, 
where \vec{E} and \vec{B} are the electric and magnetic fields in the `observers frame', 
and \vec{\beta} $=$ \vec{V} $/c$, 
then the external electric field in the flow frame disappears.
In this case, we can use the argument by  Bj\"{o}rnson (1982) that 
`the observed position angles' $\xi$ 
in the flow frame and the observer frame are the same (see Fig.~\ref{flow_obs}).
The Stokes parameters for the observer follow 
(Blandford \& K\"{o}nigl 1979, Bj\"{o}rnson 1982, Lyutikov et al. 2003)
\begin{eqnarray}
\left(
\begin{array}{c}
d I_\omega / d s \\
d Q_\omega / d s \\
d U_\omega / d s \\
d V_\omega / d s
\end{array}
\right)
= 
{\mathcal D}^2
\left(
\begin{array}{c}
j^\prime_{tot} \\
- \cos{2 (\chi_0 - \xi)} ~ j^\prime_{pol} \\
- \sin{2 (\chi_0 - \xi)} ~ j^\prime_{pol} \\
0
\end{array}
\right) \label{Stok}, 
\end{eqnarray}
where $\Gamma_f = (1-\beta^2)^{-1/2}$ and ${\mathcal D} = \Gamma^{-1}_f (1-\beta \cos{\alpha})^{-1}$ 
are the Lorentz factor and the beam factor of the flow, 
respectively, 
$\alpha$ is the angle between the direction of the observer \vec{n} 
and the flow velocity \vec{V}, $\chi_0$ is the offset caused by the practical position angle which is measured from the north 
on the sky for a given observer, 
and $\omega^\prime = \omega / {\mathcal D}$.
Numerical integration along the line of sight will be done by using grid cells constracted in the observer's frame (see Section 2.2).
Quantities in (3) such as ${\mathcal D}$, $\chi_0$, $\xi$ and $\omega^\prime$ are 
different in each grid cell because flow velocities are different cell by cell.
Suitable transformations are applied for each cell.
To each cell, the orthogonal coordinates are constructed by the unit vectors
$ \hat{\vec{V}} \equiv \vec{V}/|\vec{V}|$ in the $x$-direction, 
$ \vec{n}^\prime \times \hat{\vec{V}}$ in the $y$-direction and 
$ \hat{\vec{V}} $ $\times$ ( $ \vec{n}^\prime \times \hat{\vec{V}} $ ) in the $z$-direction, respectively.
The observed position angle $\xi$ is measured from the projected $z$-axis on the sky, i.e., 
\begin{eqnarray}
\tan{\xi} = - \frac{B^\prime_{z}}{B^\prime_{y}} \cos{\alpha^\prime} + \frac{B^\prime_{x}}{B^\prime_{y}} \sin{\alpha^\prime} ,
\end{eqnarray}
where $B^\prime_x$, $B^\prime_y$, $B^\prime_z$ are the magnetic field components in the flow frame. 
The offset angle $\chi_0$ is defined by the equation $\chi_0 \equiv -\cos^{-1}{(\hat{\mathbf Z} \cdot \hat{\mathbf z}^{\prime\prime})}$, 
where $\hat{\mathbf z}^{\prime\prime}=-\sin{\alpha} \hat{\mathbf V} + \cos{\alpha} \hat{\mathbf z}$ and $\hat{\mathbf Z}$ 
is the unit vector of the north direction on the sky. 
In Paper I, we used the transformation of the volume emissivity with the factor $\Gamma_f {\mathcal D}^3$, 
and this is not correct. 
The factor ${\mathcal D}^2$ should be used. 
Fig. 2(bottom) of Paper I may be replaced by Fig.~\ref{resultKC}(a) of the present paper. 
However, there is no apparent difference between them; the lip shape persists.

\subsection{Construction of polarization map}
\label{const}
If the nebula is optically thin, it is straightforward to obtain
the observed Stokes parameters, simply by integration of (\ref{Stok}) along the line of sight.
We use the observer's frame ($X$,$Y$,$Z$) such that 
the observer locates at $X=+ \infty$, 
the $YZ$ plane defines the sky with the north in $Z$-direction, 
and the nebula center (pulsar) is located at the origin.
The volume emissivities of (\ref{Stok}) are functions of $X$,$Y$,$Z$ and \vec{n}, 
so that maps of the stokes parameters are given by 
\begin{eqnarray}
\begin{array}{l}
I_{\omega} (Y,Z) = \displaystyle{\int^{+\infty}_{-\infty} \frac{dI_{\omega} (X,Y,Z,\vec{n})}{dX} dX},\\ 
Q_{\omega} (Y,Z) = \displaystyle{\int^{+\infty}_{-\infty} \frac{dQ_{\omega} (X,Y,Z,\vec{n})}{dX} dX},\\
U_{\omega} (Y,Z) = \displaystyle{\int^{+\infty}_{-\infty} \frac{dU_{\omega} (X,Y,Z,\vec{n})}{dX} dX}. 
\end{array}\label{INT}
\end{eqnarray}
We construct grid points on the plane of sky ($Y_i$,$Z_j$) for polarization maps, 
and on the line of sight $X_k$ for integration (\ref{INT}) with increment of $\Delta s$. 
The Stokes parameters as the whole nebula are obtained by 
\begin{eqnarray}
\begin{array}{l}
I_{t} = \displaystyle \sum_{i} \displaystyle \sum_{j} I(Y_i,Z_j),\\
Q_{t} = \displaystyle \sum_{i} \displaystyle \sum_{j} Q(Y_i,Z_j),\\
U_{t} = \displaystyle \sum_{i} \displaystyle \sum_{j} U(Y_i,Z_j),
\end{array}
\end{eqnarray}
with which the mean polarization degree $\bar{P}$ and position angle (P.A.), $\bar{\chi}$, 
are obtained by 
$\bar{P} = \sqrt{{Q_t}^2 + {U_t}^2}/I_t$ and 
$\bar{\chi} = (1/2) \tan^{-1} \left( U_t/Q_t \right)$, respectively.

\subsection{Crab Nebula model}
\label{model}
We start our calculation with a simple disc model with pure toroidal magnetic field (Paper I).
The disc represents the post shock flow with an inner radius $R_s$ located at the shock 
and with a constant semi-opening angle $\theta_0$. 
The flow is based on KC: 
the radial flow velocity $V(R)$ is given (KC), 
where $R=(X^2+Y^2+Z^2)^{1/2}$. 
The post shock flow is characterized 
by the $\sigma$ parameter which is the ratio of Poynting flux 
to the kinetic energy flux just before the shock. 
The nebula flow suffers adiabatic and synchrotron losses.
In Paper I, 
we calculate the evolution of distribution function. 
The parameters $K$ and $p$ in (\ref{Stok_0}) 
are determined as functions of $R$ 
so as to fit the distribution function obtained 
in Paper I at each point in the nebula. 
The magnetic field distribution $B_{KC} (R)$ is also given by the KC model.
We use following a parameter set: 
$\sigma=0.003$, 
the wind luminosity $L_w=5 \times 10^{38}$ erg~s$^{-1}$, 
the wind Lorentz factor $\gamma_w = 3 \times 10^6$, 
the shock distance $R_s = 3 \times 10^{17}$ cm, 
the power-law index at the shock $p_s = 3$, 
and the thickness of the disc $\theta_0 = \pm 10^\circ$ degree. 
The inclination angle of the axis of the disc to the observer is $i = 28^\circ$ (Weisskopf et al. 2000).

\subsection{The case of disordered toroidal field}
\label{disB}
It has been suggested in Paper I that the nebula field is not 
`pure toroidal' but may be dominated by disordered fields. 
Scale length of the randomness (turbulent spectrum) is not known; 
it can be microscopic or just below the resolution of observation.
In this section, let us consider the case of disordered magnetic field. 

In general, the magnetic field can be decomposed to the mean field $\vec{B}^\prime_0$ 
and the random field $\vec{B}^\prime_1$. 
The degree of randomness may be characterized by 
\begin{eqnarray}
b= \frac{<{\vec{B}^\prime_1}^2>}{{B^\prime_0}^2+<{\vec{B}^\prime_1}^2>}, 
\end{eqnarray}
where $< >$ indicates the spatial average in a scale below the observational resolution 
and the primes indicate that $b$ is evaluated in the flow frame ($<{\vec{B}^\prime_1}^2>$ is still a function of position). 
The synchrotron radiation for such cases is studied by Korchakov \& Syrovat-skii (KS) (1962). 
From (3), including relativistic motion of the flow, 
we have the observed Stokes parameters: 
\begin{eqnarray}
\left(
\begin{array}{c}
d I_\omega / d s \\
d Q_\omega / d s \\
d U_\omega / d s \\
d V_\omega / d s
\end{array}
\right)
= 
{\mathcal D}^2
\left(
\begin{array}{c}
j^\prime_{tot} \\
- \cos{2 (\chi_0 - \xi_0)} ~ j^\prime_{pol} \\
- \sin{2 (\chi_0 - \xi_0)} ~ j^\prime_{pol} \\
0
\end{array}
\right), \label{Stok_random} 
\end{eqnarray}
where 
\begin{eqnarray}
\begin{array}{l}
j^\prime_{tot} = \displaystyle{\frac{p+7/3}{p+1} \Phi(\omega^\prime, p)} \\
\quad \times \displaystyle{\frac{1}{\Delta s} \int^{s+ \Delta s}_{s} \textstyle{\left( B^\prime \sqrt{1-\left(\vec{n}^\prime \cdot \frac{\vec{B}^\prime}{\displaystyle{B^\prime}}\right)^2} \right)}^{\frac{p+1}{2}} ds^\prime}, \label{j_prime_tot_random}\\
\end{array}
\end{eqnarray}
\begin{eqnarray}
\begin{array}{l}
j^\prime_{pol} = \displaystyle{\Phi(\omega^\prime, p)} \\
\quad \times \displaystyle{\frac{1}{\Delta s} \int^{s+ \Delta s}_{s} \textstyle{\left( B^\prime \sqrt{1-\left(\vec{n}^\prime \cdot \frac{\vec{B}^\prime}{\displaystyle{B^\prime}}\right)^2} \right)}^{\frac{p+1}{2}} \cos{2 \chi} ~ds^\prime}, \label{j_prime_pol_random}\\
\end{array}
\end{eqnarray}
\begin{eqnarray}
\begin{array}{l}
B^\prime = \displaystyle{\sqrt{ | \vec{B}^\prime |^2} = \sqrt{ {B^\prime_0}^2 + 2 \vec{B}^\prime_{0} \cdot \vec{B}^\prime_{1} + {B^\prime_{1}}^2 } }, 
\end{array}
\end{eqnarray}
\begin{eqnarray}
\begin{array}{l}
\cos{2 \chi} = \displaystyle{2 \left( \frac{ \vec{B}^\prime_{0} \times \vec{n}^\prime}{|\vec{B}^\prime_0 \times \vec{n}^\prime|} \cdot \frac{ \vec{B}^\prime \times \vec{n}^\prime}{|\vec{B}^\prime \times \vec{n}^\prime|} \right)^2 -1 } \label{cos2chi}.
\end{array}
\end{eqnarray}
The volume emissivities (\ref{Stok_random}) are obtained for each grid cell 
and are integrated numerically according to (\ref{INT}). 
Given by a nebula model, 
macroscopic quantities such as $\vec{V}$ and $\vec{B}_0$ 
are assumed to be constant in a given grid cell, 
but gradually change cell by cell. 
In each cell, averaging for the random field (\ref{j_prime_tot_random}) and (\ref{j_prime_pol_random}) is done by Monte Carlo method. 
KS use the special coordinate in which $U$ vanishes. 
By using this coordinate, only $Q$ remains, 
and Monte Carlo integration can be only one time; 
the factor $\cos{2 \chi}$ appears in (\ref{j_prime_pol_random}), 
but does not $\sin{2 \chi}$, 
where $\chi$ is defined by (\ref{cos2chi}) and represents the angle 
between the polarization E-vector and the projected magnetic field. 
The value of $\chi$ varies place by place within the cells. 
Redistribution of the Stokes parameters for the observers coordinate can be done by (\ref{Stok_random}) 
by using the angle $\xi_0$ defined by the projected mean field: 
\begin{eqnarray}
\tan{\xi_0} = - \frac{B^\prime_{0z}}{B^\prime_{0y}} \cos{\alpha^\prime} + \frac{B^\prime_{0x}}{B^\prime_{0y}} \sin{\alpha^\prime}. 
\end{eqnarray}

As stated above, 
integration along the line of sight has been done in use of Monte Carlo method with randomly distributed $\vec{B}^\prime_1$'s. 
In the present calculation, 
we assume for comparison purposes that the energy density 
of the nebula field is equal to the KC value $B^2_{KC}$, 
i.e., 
${B^\prime_0}^2 + <{\vec{B}^\prime_1}^2> = B^2_{KC}/\Gamma^2_f$, 
and that the plasma's distribution functions are the same as Paper I.
It is obvious that the disordered fields change flow dynamics: 
(i) the disordered fields contribute to the magnetic pressure rather than magnetic tension, 
(ii) formation of the disordered fields must be associated by heating, 
and (iii) the distribution function must be changed. 
These issues are postponed to subsequent papers.

Apart from the disordered field, it is also suggested by 
some authors (Hester et al. 2002, Mori et al. 2004, Ng \& Romani 2004, and Paper I) 
that if the front-back (north-west to south-east) 
intensity contrast is due to Doppler boost, 
the flow velocity is much faster than the KC flow ($\sim 2000 \ {\rm km} \; {\rm s}^{-1}$) and 
will be $\sim 0.2 c$. 
Therefore, we also calculate the case with much faster flows. 

In summary, we calculate three cases; 
Case KC, in which plasma properties follows the KC model (see Section 2.3 in detail); 
Case D, in which we include the disordered field with $b$ (the degree of randomness). 
For comparison purposes the total magnetic energy density and 
flow speed are left unchanged; 
Case DR, in which, in addition to Case D 
we change the flow velocity to be $0.2 c$ through out the whole nebula 
by hand (we ignore the flow dynamics).

\section{Results}
\subsection{Case of Pure Toroidal Field}
\label{RuniB}
\begin{figure}
\begin{center}
 \begin{minipage}{0.48\linewidth}
   \includegraphics[angle=-90,width=\linewidth]{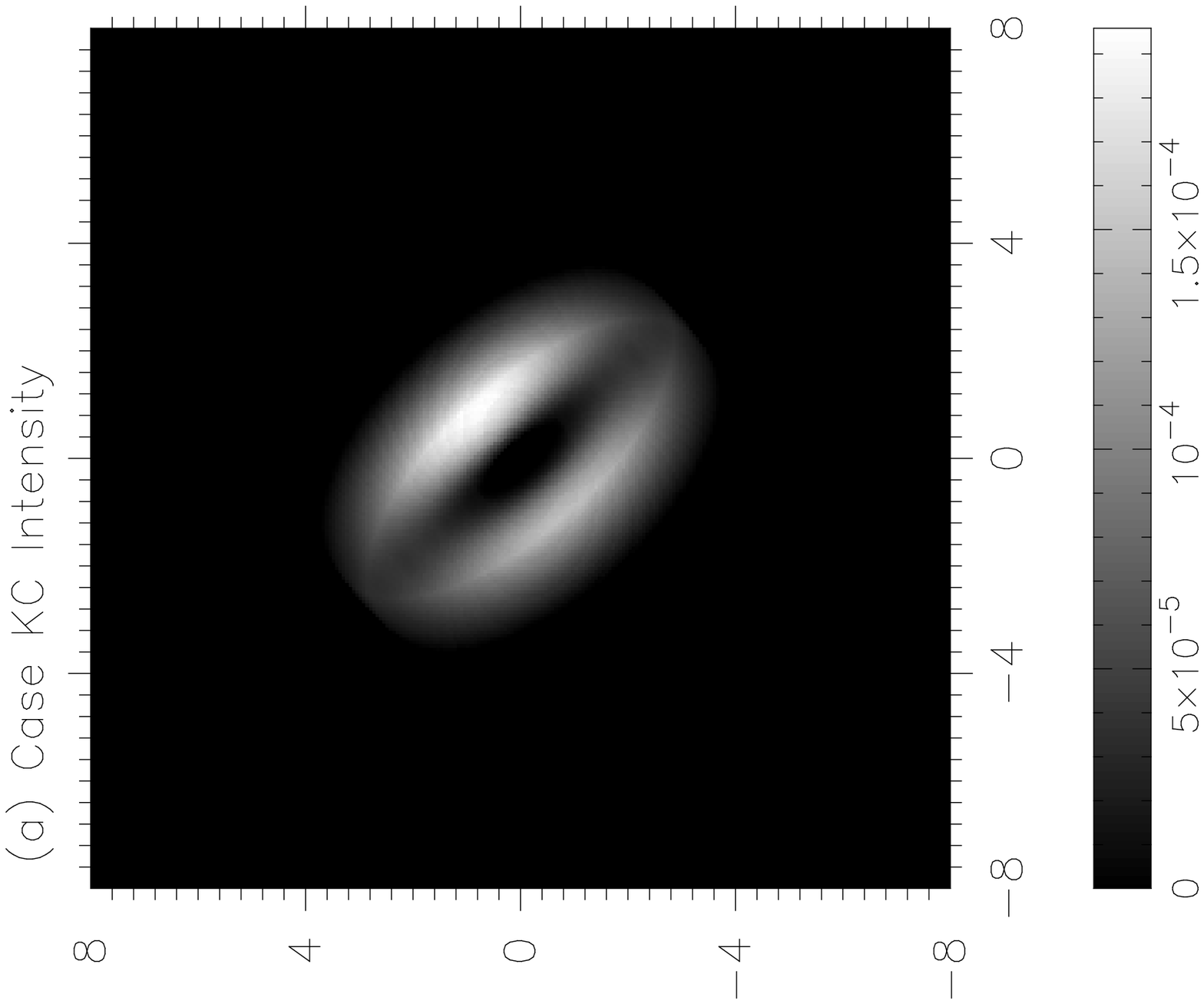}
 \end{minipage}
 \begin{minipage}{0.48\linewidth}
   \includegraphics[angle=-90,width=\linewidth]{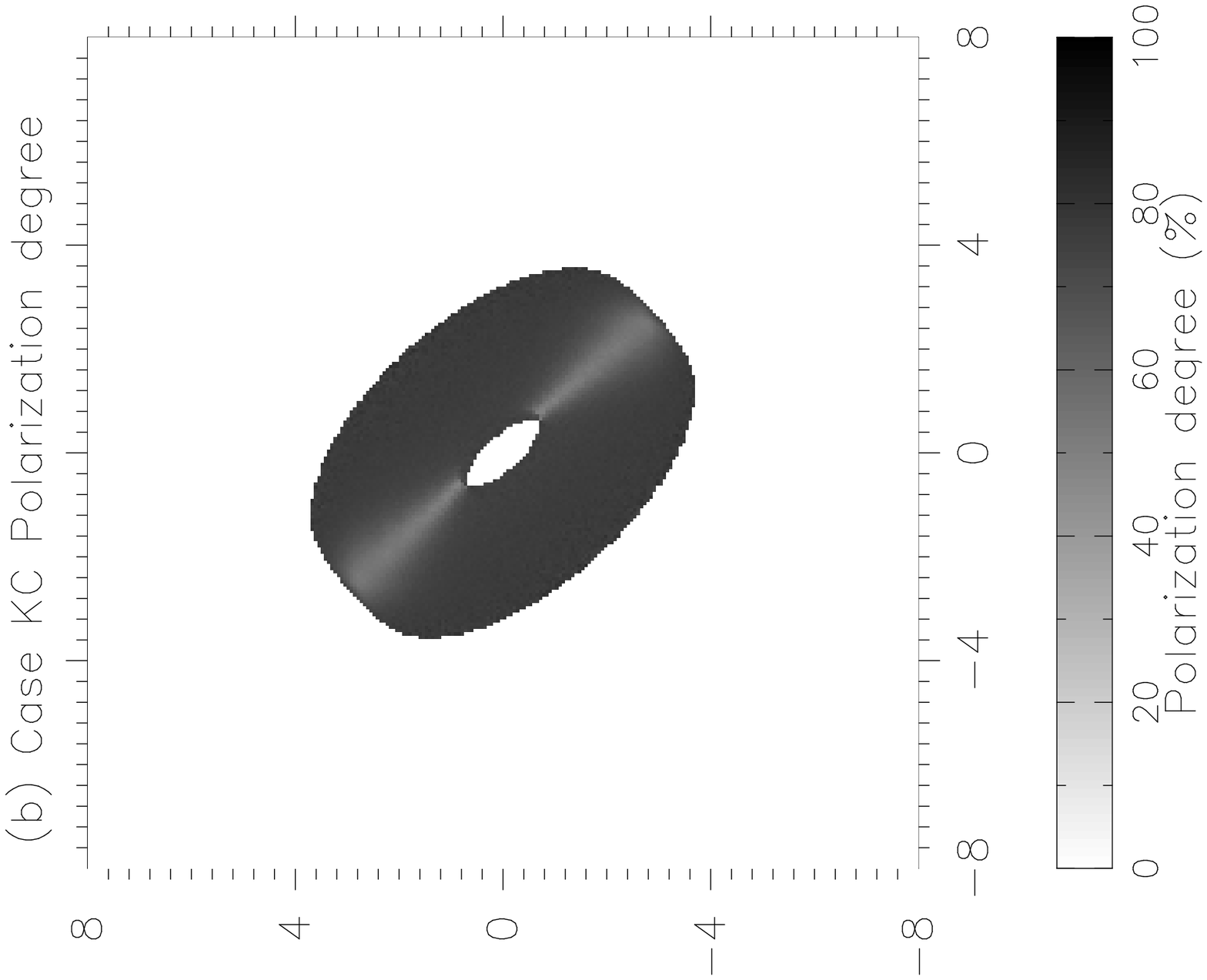}
 \end{minipage}
\\[5mm]
 \begin{minipage}{0.48\linewidth}
   \includegraphics[angle=-90,width=\linewidth]{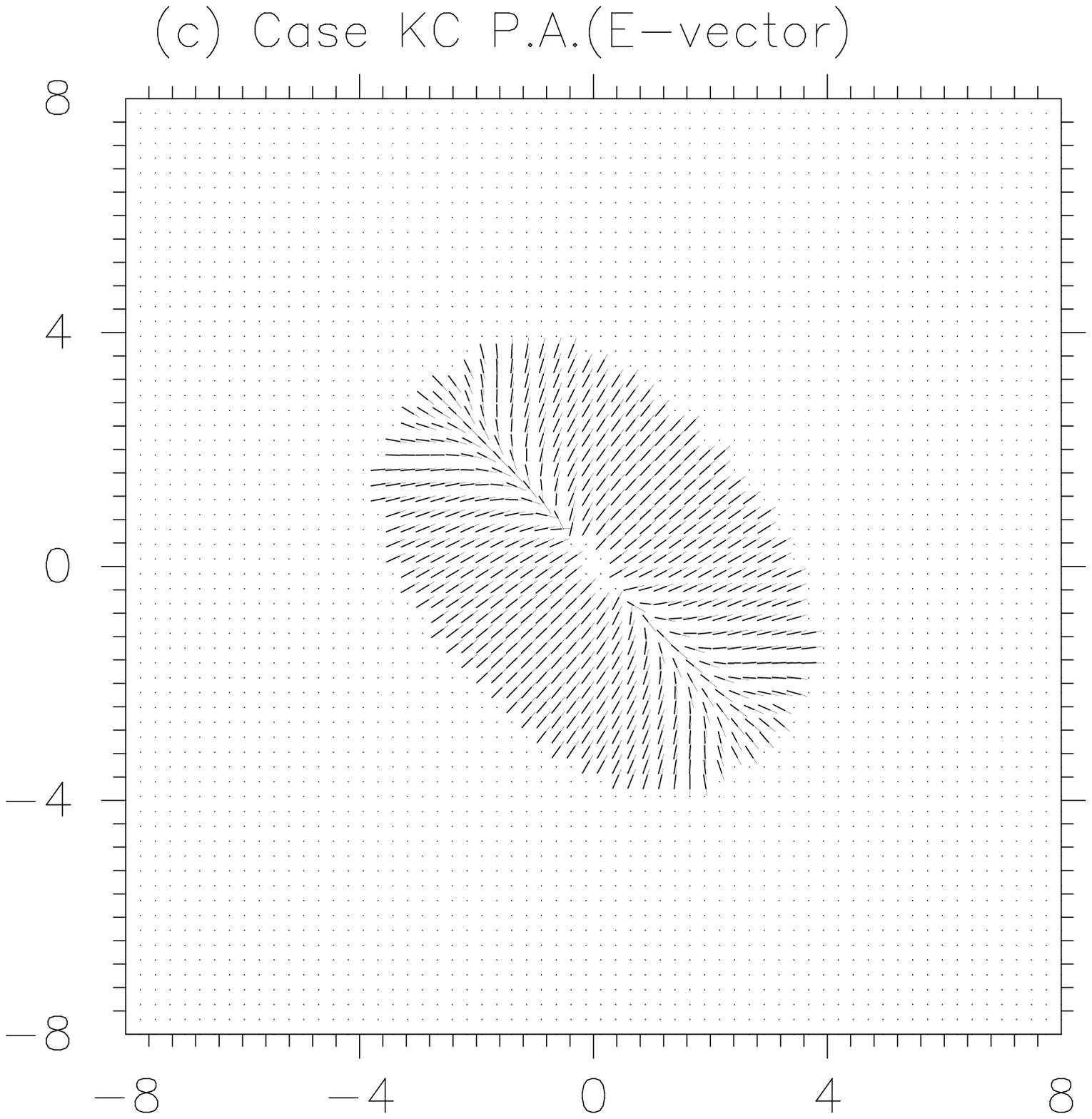}
 \end{minipage}
 \begin{minipage}{0.48\linewidth}
   \includegraphics[width=\linewidth]{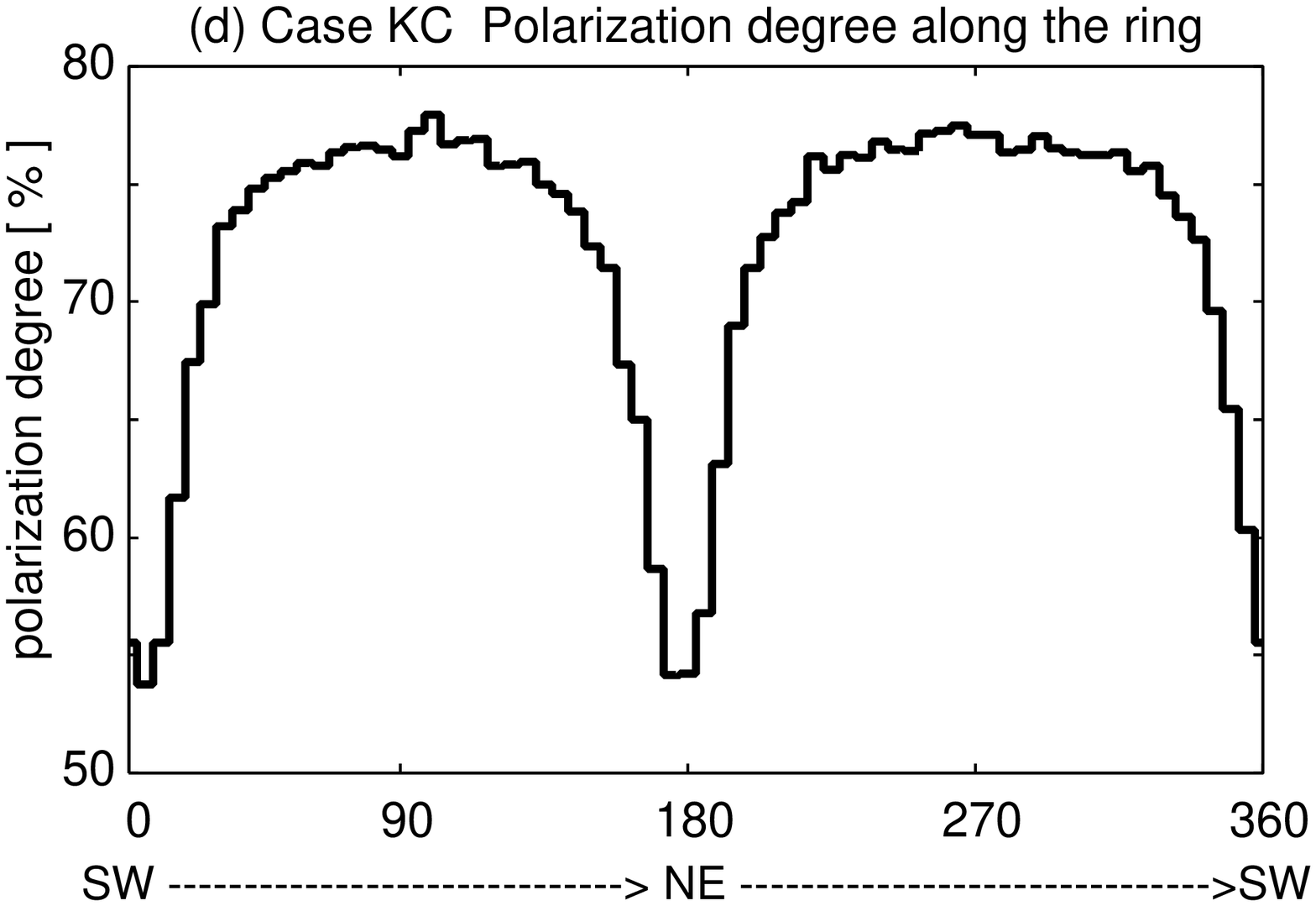}
 \end{minipage}
\end{center}
\caption{ \label{resultKC}
Polarization properties of the pulsar disc nebula for Case KC, 
observed at $5.2$ keV.
(a)Intensity map. 
The unit length is a shock radius and the unit of the intensity is 0.016 erg~s$^{-1}$~cm$^{-2}$~str$^{-1}$~eV$^{-1}$. 
(b)Polarization degree map 
drawn only for the regions where the intensity is strong enough, regarding calculation errors.
(c)P.A. map (E-vector). 
(d)Polarization degree as a function 
of the azimuthal angle along an ellipse with the 
semi-major axis of $2.5 R_s$ and semi-minor axis of $1.5 R_s$.
The labels `SW', `NE' and `SW' below the horizontal axis 
indicate south-west, north-east and south-west, respectively.
}
\end{figure}
Fig.~\ref{resultKC} shows the results of Case KC.  
The observation energy is assumed to be $5.2$ keV.
The intensity map (Fig.~\ref{resultKC}(a)) 
shows a lip-shaped nebula, 
as was pointed out in Paper I.
On the major axis (NE-SW) of the ring (perpendicular to the rotation axis of the pulsar) and its vicinity, 
the pitch angles of the particles radiating toward the observer are small, 
and as a result synchrotron emissivity is decreased. 
The dimness is cased not only by decrease of single-particle emissivity 
but also by decrease of number of the particles emitting to the observing bands. 
Fig.~\ref{resultKC}(b) shows the polarization degree. 
In Fig.~\ref{resultKC}(d), the polarization degree 
is plotted as an azimuthal function along the ring. 
On the minor axis (rotational axis of the pulsar), 
the polarization degree reaches the maximum value of $\sim  75\% $.
There is depolarized regions along the major axis, 
where the polarization degree decreases to $\sim 55\%$. 

In Case KC, 
the mean polarization degree as the whole nebula is $52.3 \%$, 
and P.A. is found to be $132^\circ$ at $5.2$ keV. 
This polarization degree is much larger than the observed value ($\sim 20 \%$) in X-ray (Weisskopf et al. 1978).
Although we calculate also for optical bands, 
the results are hardly changed so that they are not shown in figures.
Since the mean polarization in optical is $\sim 19 \%$ (Oort and Walraven 1956), 
the calculated polarization degree in optical is again much higher than the observed. 
\begin{figure}
\begin{center}
\includegraphics[clip,width=8.4cm]{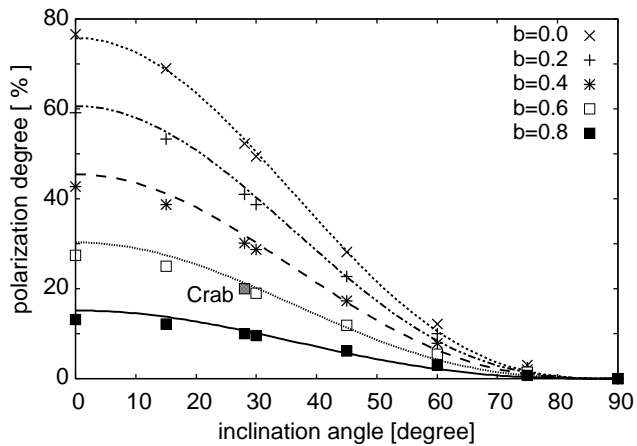} 
\end{center}
\caption{ \label{Pofbi}
Polarization of the entire nebula as functions 
of the inclination angle $i$ between the axis of the disc and the line of sight for $b=0.0$, $0.2$, $0.4$, $0.6$ and $0.8$.
The empirical curves $\bar{P} = 75.8 (1-b) (\cos{i})^{2.84}$ are overlaid.
}
\end{figure}

\subsection{Effects of Disordered Field and Relativistic Motion}
\label{RdisB}
In Case D, 
we change the parameter $b$ to find the value which reproduces the observed polarization degree.
Fig.~\ref{Pofbi} shows how the mean polarization behaves 
as a function of $b$ and the inclination angle $i$, 
the angle of the symmetry axis of the nebula from \vec{n} (the line of sight). 
The result can be fitted by an empirical relation, 
$\bar{P} = 75.8 \times (1-b) \times (\cos{i})^{2.84}$. 
For the Crab Nebula, the inclination angle is known to be $i = 28^\circ$ (Weisskopf et al. 2000), 
and therefore the observed polarization degree can be reproduced if $b=0.6$. 

\begin{figure}
\begin{center}
 \begin{minipage}{0.48\linewidth}
   \includegraphics[angle=-90,width=\linewidth]{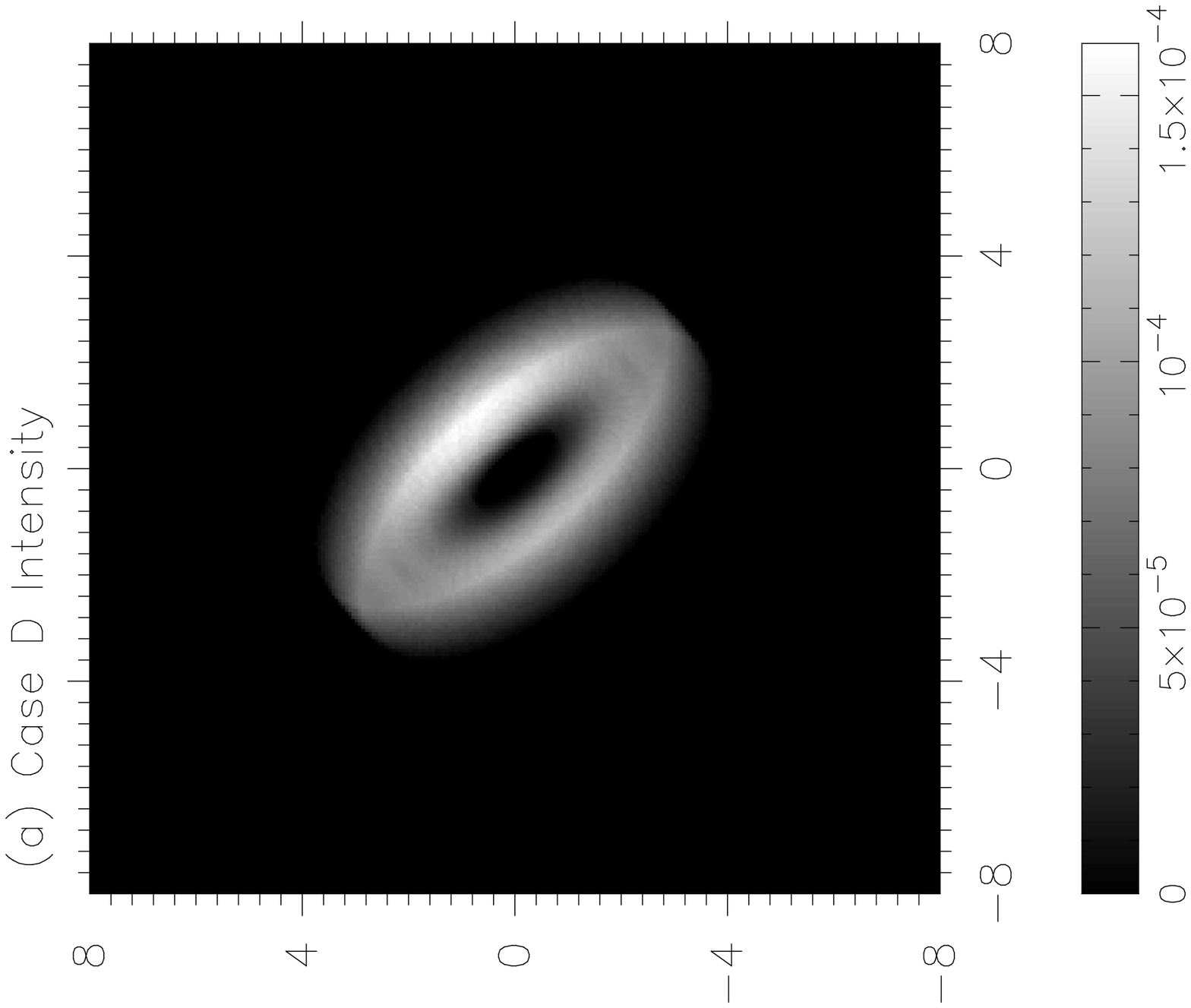}
 \end{minipage}
 \begin{minipage}{0.48\linewidth}
   \includegraphics[angle=-90,width=\linewidth]{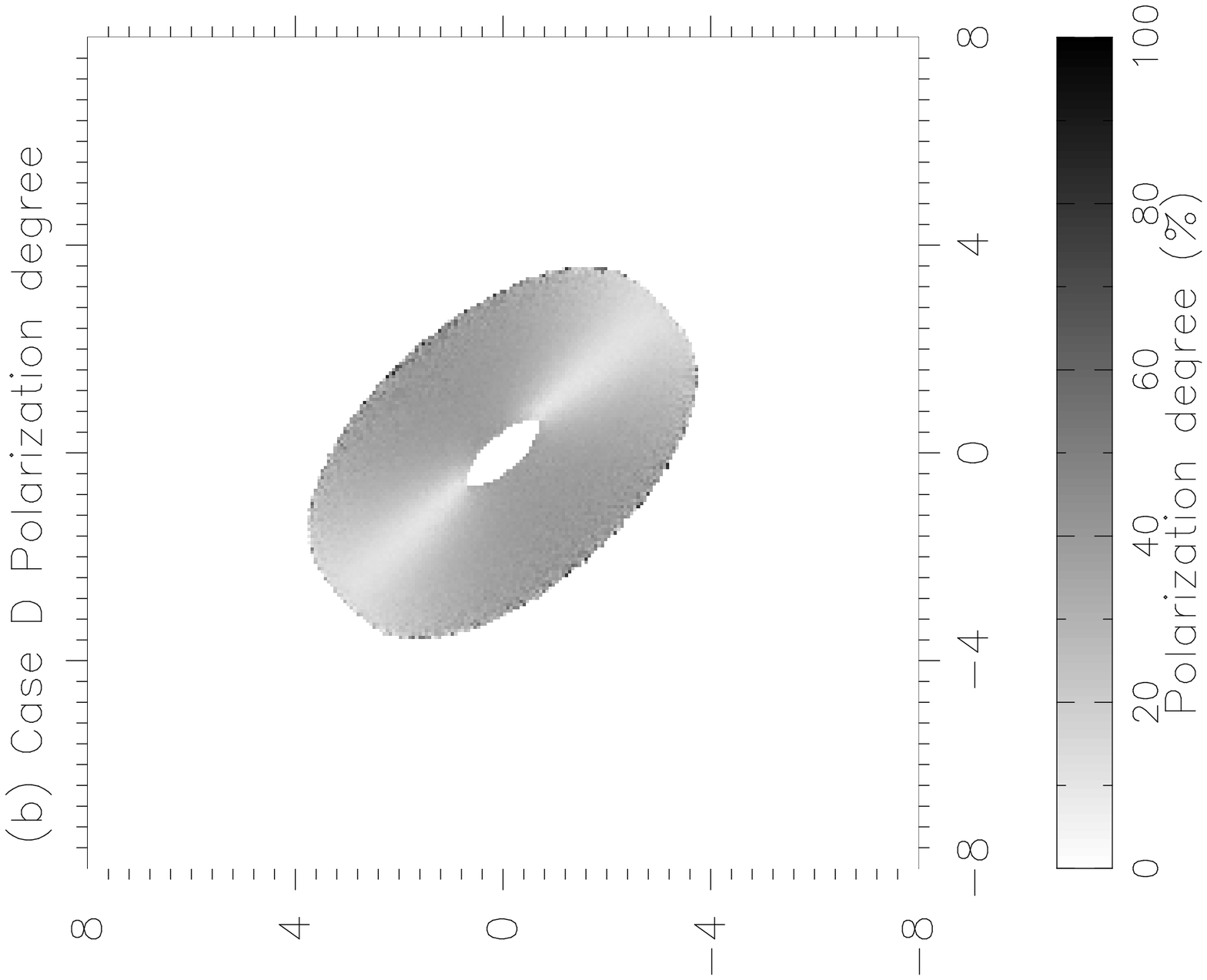}
 \end{minipage}
\\[5mm]
 \begin{minipage}{0.48\linewidth}
   \includegraphics[angle=-90,width=\linewidth]{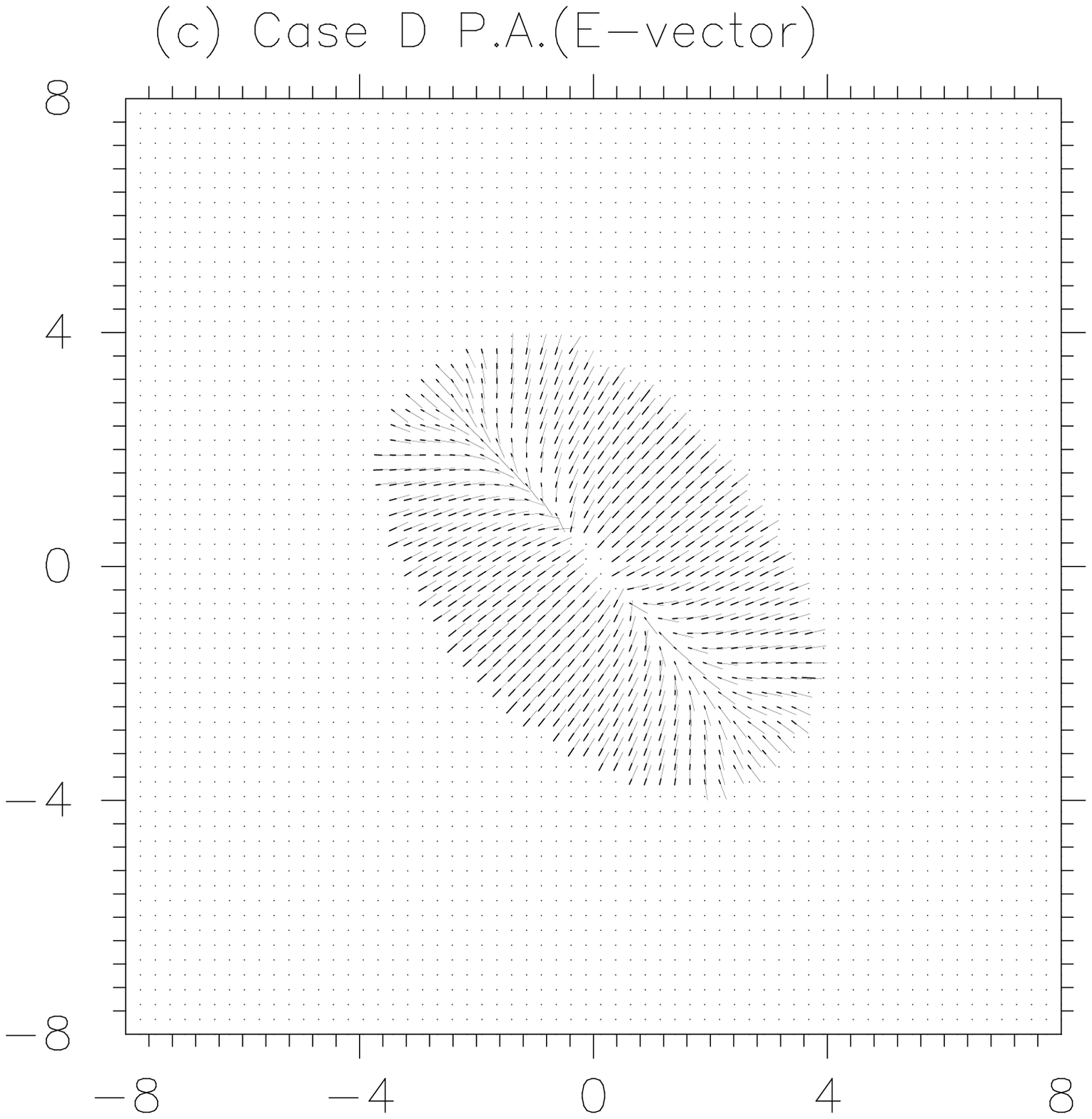}
 \end{minipage}
 \begin{minipage}{0.48\linewidth}
   \includegraphics[width=\linewidth]{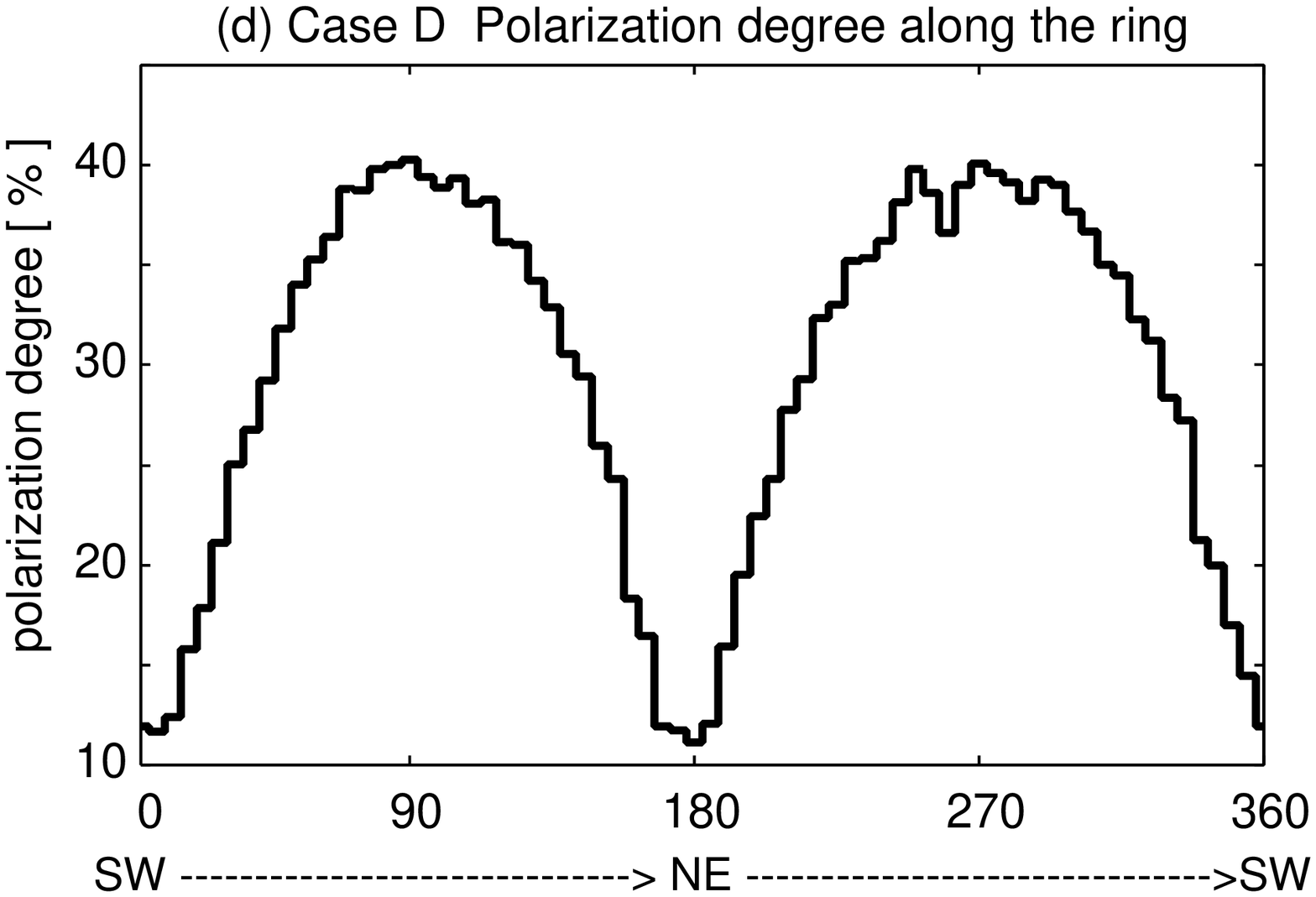}
 \end{minipage}
\end{center}
\caption{ \label{resultD}
Same as Figure \ref{resultKC} but for Case D with $b=0.6$.
}
\end{figure}

Polarization properties for Case D with $b=0.6$ are summarized 
in Fig.~\ref{resultD} at the observation energy of $5.2$ keV. 
As has been mentioned in Section 1, 
almost the same polarization degrees are observed both in X-ray and optical bands. 
The calculated polarization properties for the optical bands are hardly different from those for the $5.2$ keV band.
This is thus consistent with the observation. 
The polarization degree takes the maximum value of $\sim 40 \%$ 
on the northwest side and the southeast side of the ring.
This also agrees with the observation in optical bands 
(Schmidt et al. 1979, Hickson \& van der Bergh 1990, Michel et al. 1991). 

Case D is favored by at least two observational facts 
that (1) the intensity map appears in a ring (not lip-shaped), 
and that (2) the polarization degree is $\sim 40 \%$ 
at the highest and $\sim 20 \%$ on average. 
This indicates that the disordered magnetic component is $60 \%$ of the total.
  
\begin{figure}
\begin{center}
 \begin{minipage}{0.48\linewidth}
   \includegraphics[angle=-90,width=\linewidth]{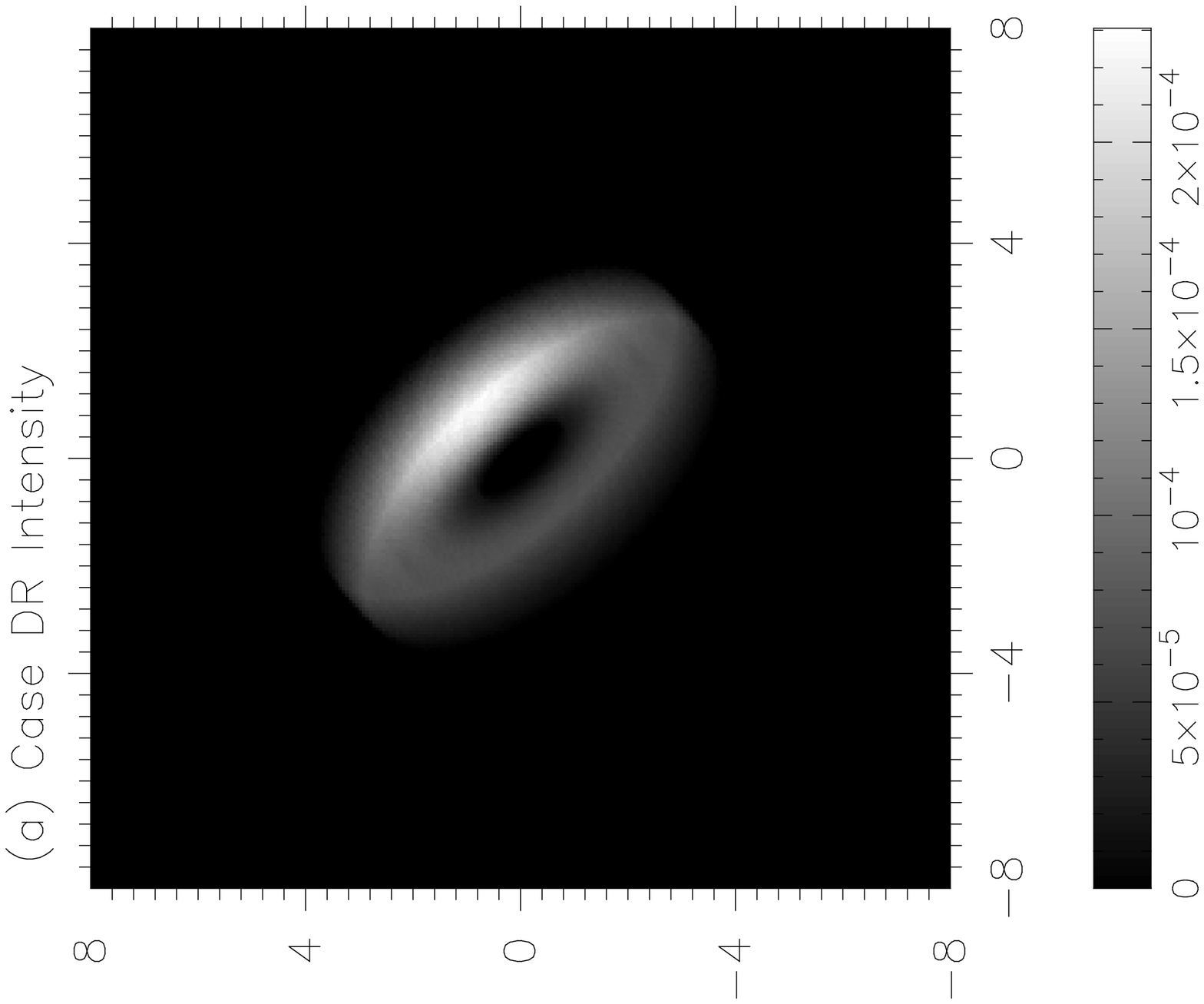}
 \end{minipage}
 \begin{minipage}{0.48\linewidth}
   \includegraphics[angle=-90,width=\linewidth]{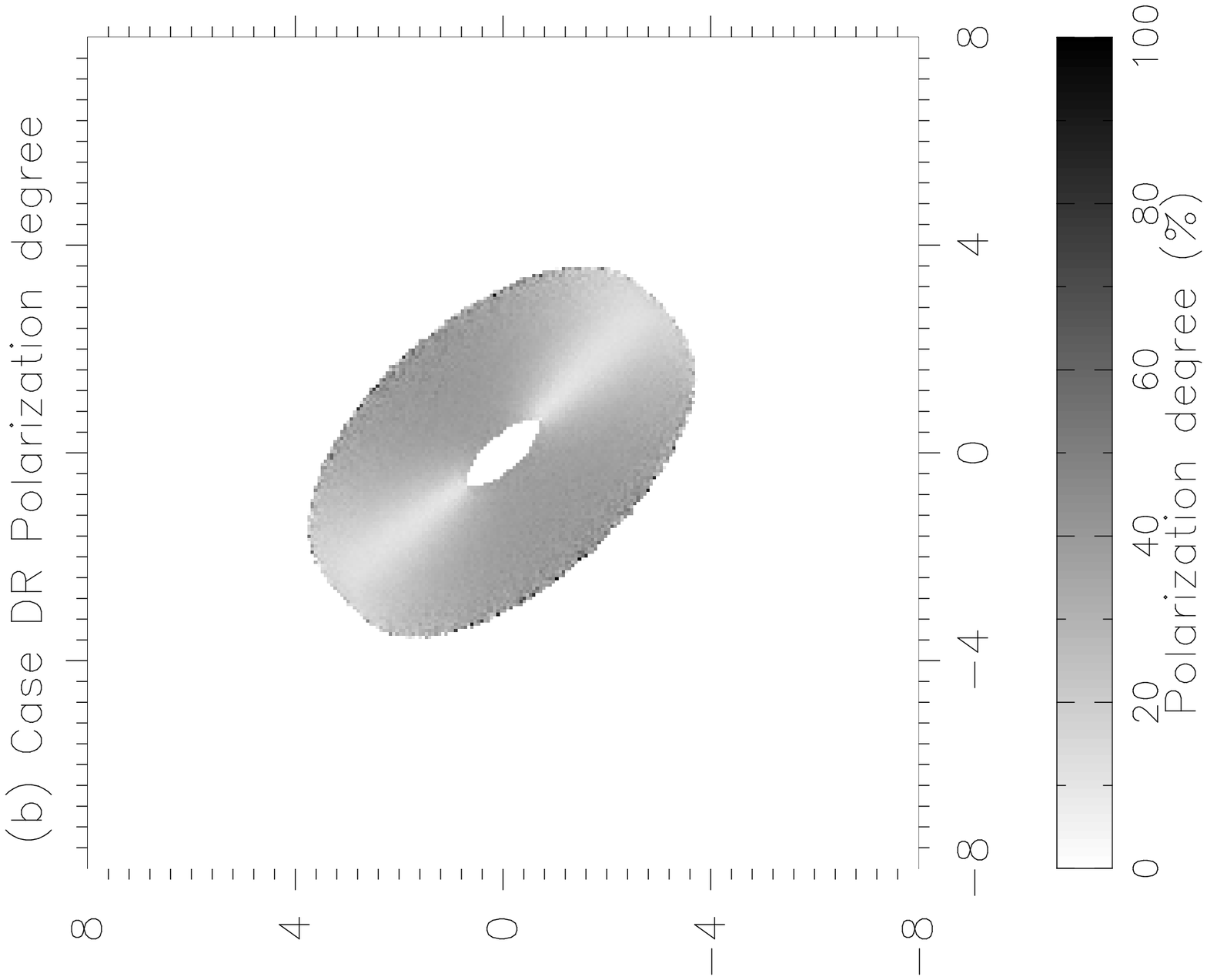}
 \end{minipage}
\\[5mm]
 \begin{minipage}{0.48\linewidth}
   \includegraphics[angle=-90,width=\linewidth]{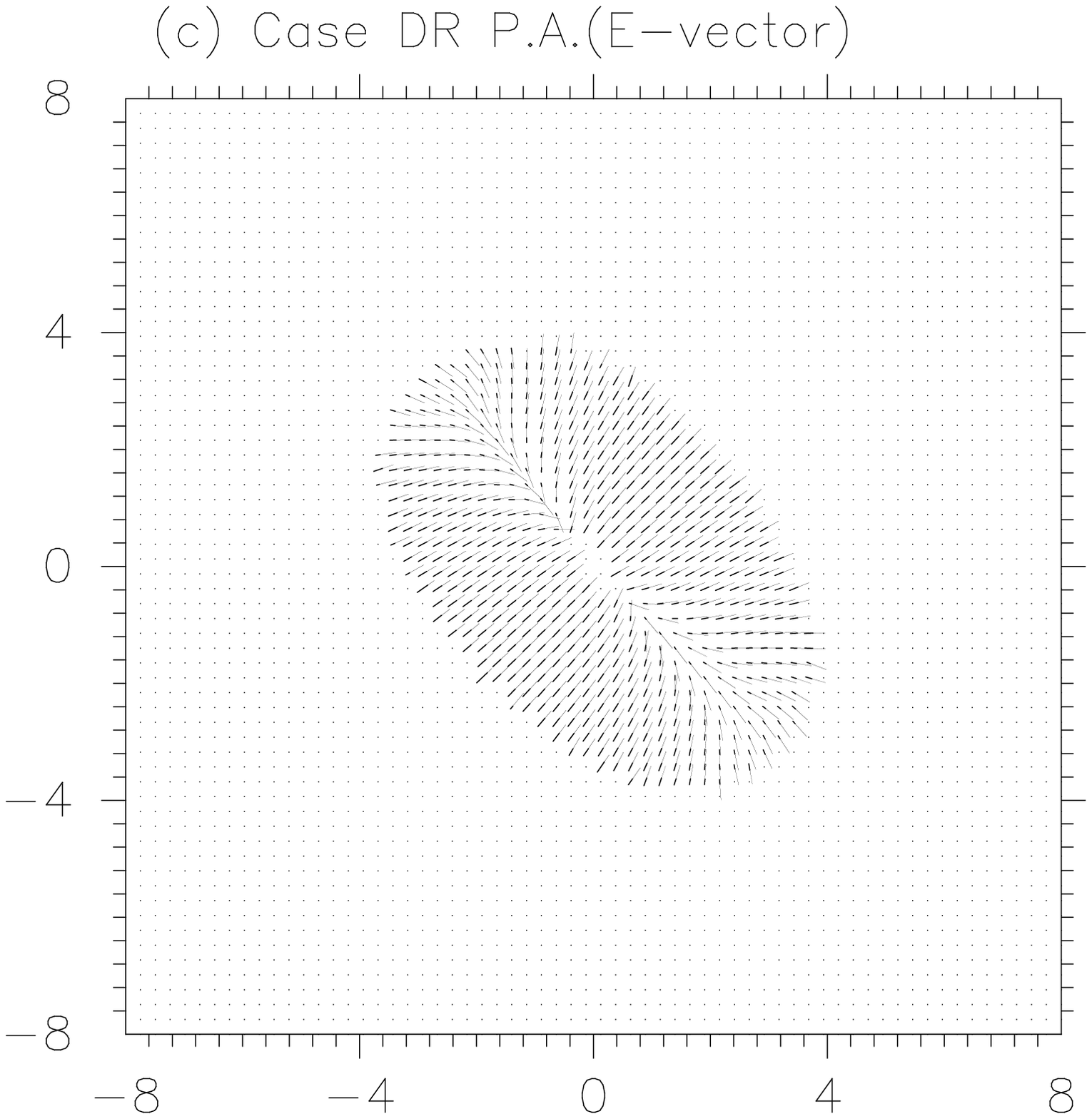}
 \end{minipage}
 \begin{minipage}{0.48\linewidth}
   \includegraphics[width=\linewidth]{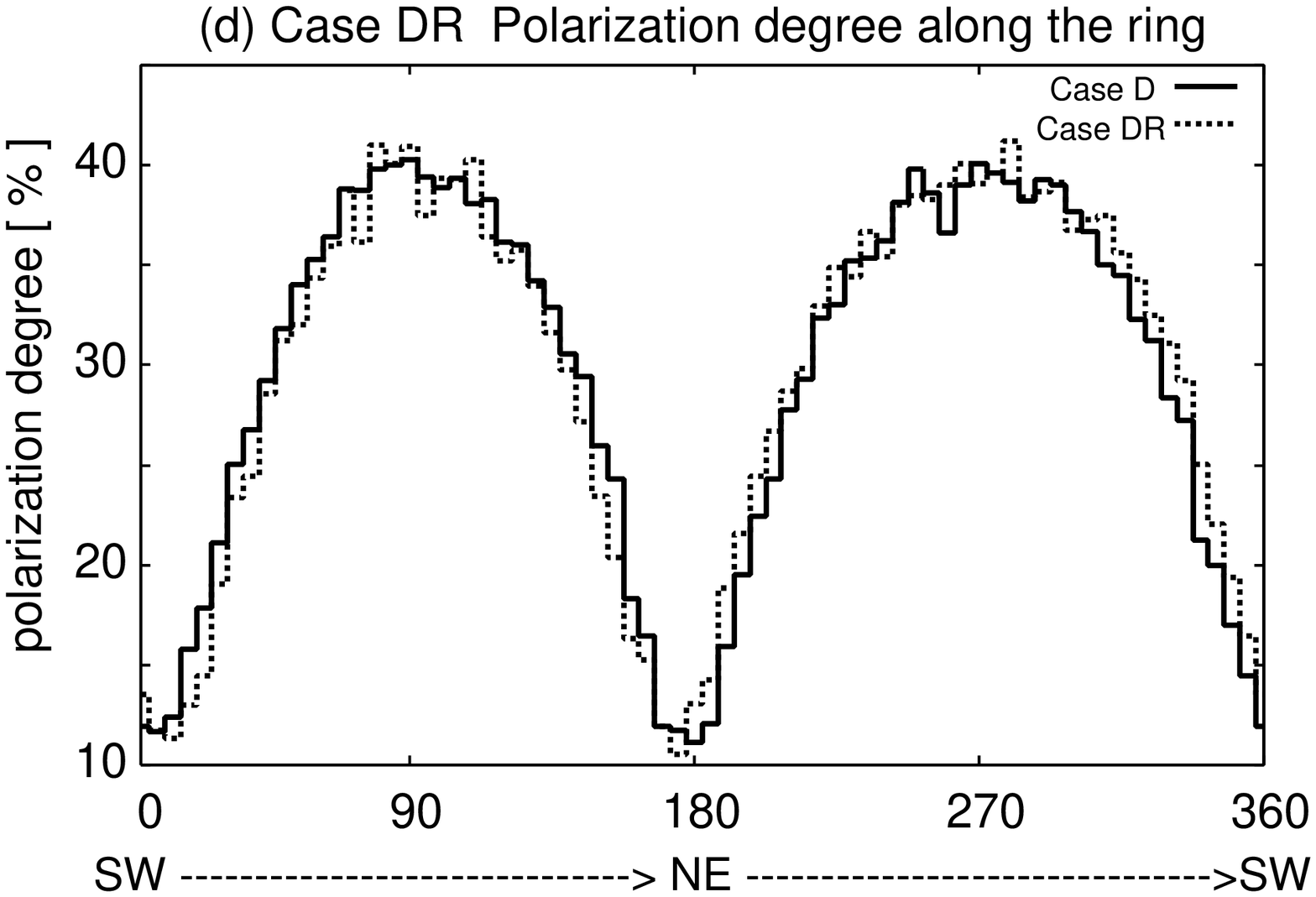}
 \end{minipage}
\end{center}
\caption{ \label{resultDR}
Same as Figure \ref{resultKC} but for Case DR with $b=0.6$, $v_r=0.2c$.
(d) shows the comparison between the polarization degree of Case D (solid line) 
and that of Case DR (dotted line).
}
\end{figure}

The front-back contrast in Case D is smaller 
than the observed value $3.4$ (Mori et al. 2004) 
because the model follows the KC flow. 
In Case DR, 
the flow velocity is assumed to be $0.2 c$ in the whole nebula, 
and thereby front-back contrast becomes $3.1$ (Fig.~\ref{resultDR}).
The average polarization of Case DR is increased only by $0.14 \%$ as compared with Case D.
This increase is caused by the high polarization degree 
at the north-west region weighted by Doppler boosting. 
Relativistic effects is rather prominent in distribution of the depolarization. 
When the nebula flow is non-relativistic as seen in Case KC and Case D, 
the depolarization takes place along the major axis symmetrically, 
but in Case DR the depolarized regions shift slightly to the northwest side 
which expands toward us (Fig.~\ref{resultDR}(b) and (d)). 
The electric field vectors of the synchrotron radiation shown in Fig.~\ref{resultDR}(c) is no longer vertical 
to the projected magnetic field on the sky as the flow becomes relativistic. 
These relativistic effect is difficult to see in the complex optical polarization maps. 
However, 
it might be verified if spatially-resolved polarization in X-ray bands is performed 
because the ring structure has a better symmetry in X-ray, 
and obscuring by filaments is less. 
After all, 
the relativistic effect does not change the estimate of the disordered magnetic field. 

\begin{figure}
\begin{center}
 \begin{minipage}{0.48\linewidth}
   \includegraphics[width=\linewidth]{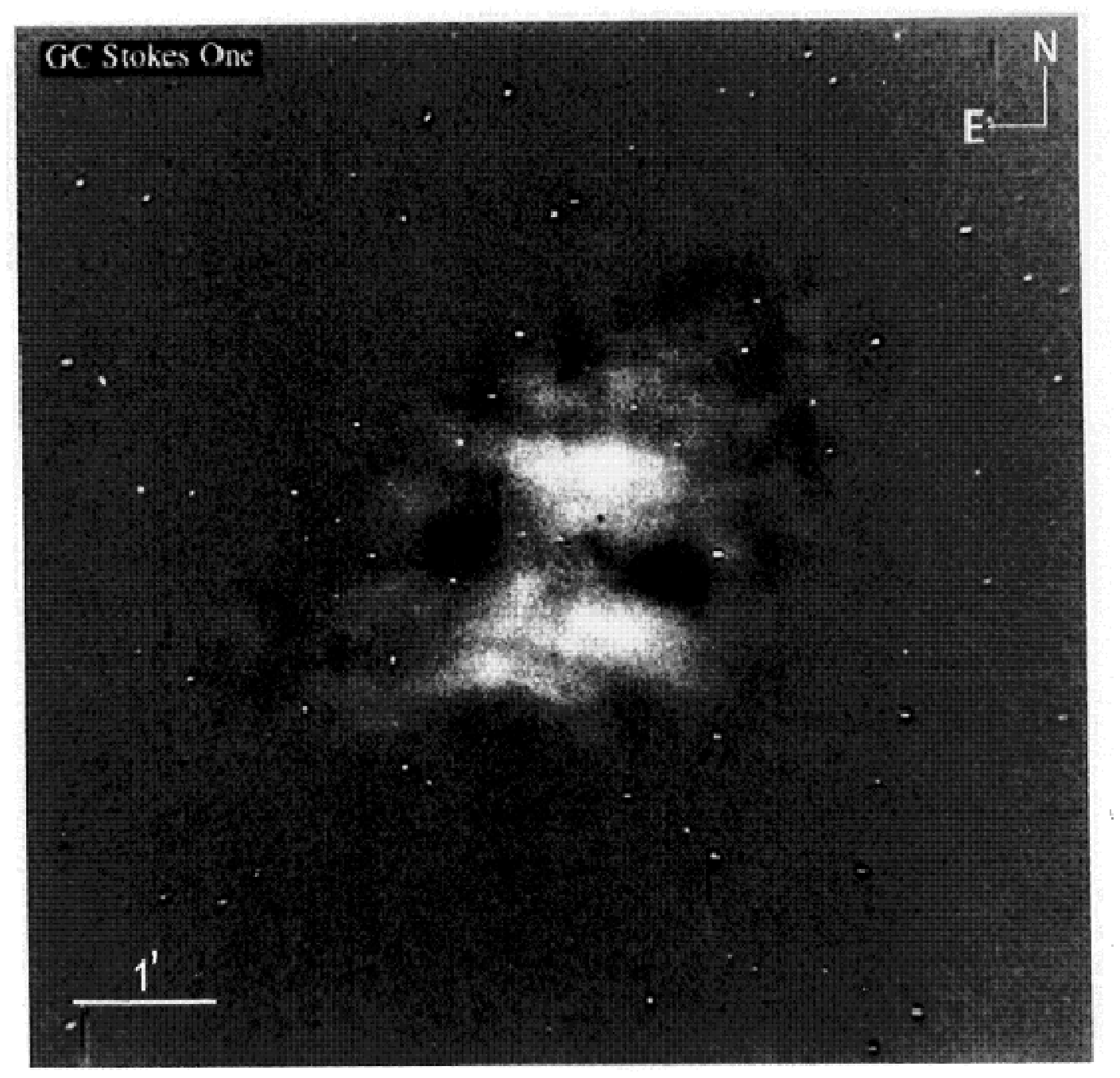}
 \end{minipage}
 \begin{minipage}{0.48\linewidth}
   \includegraphics[width=\linewidth]{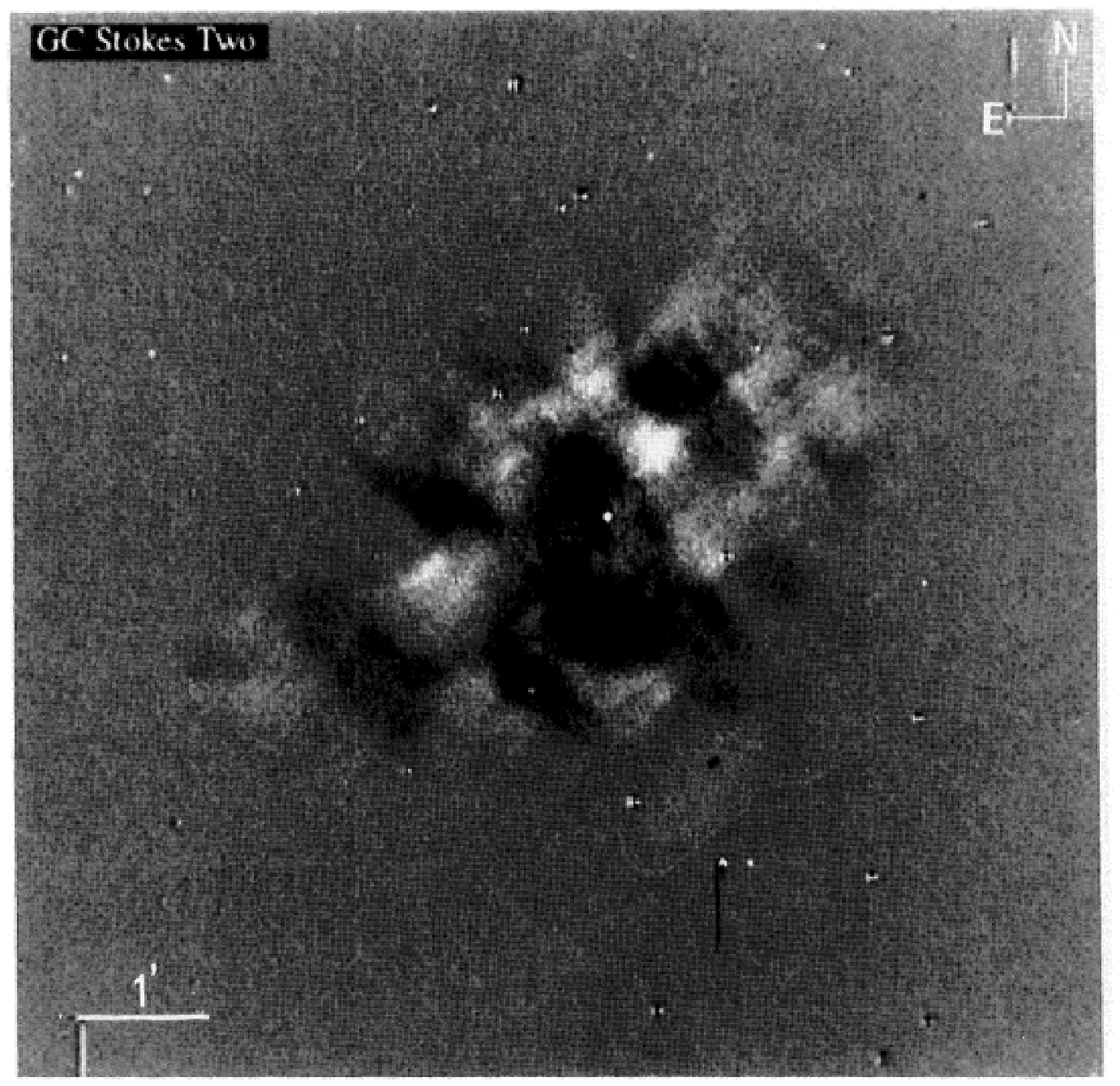}
 \end{minipage}
\\[5mm]
 \begin{minipage}{0.48\linewidth}
   \includegraphics[angle=-90,width=\linewidth]{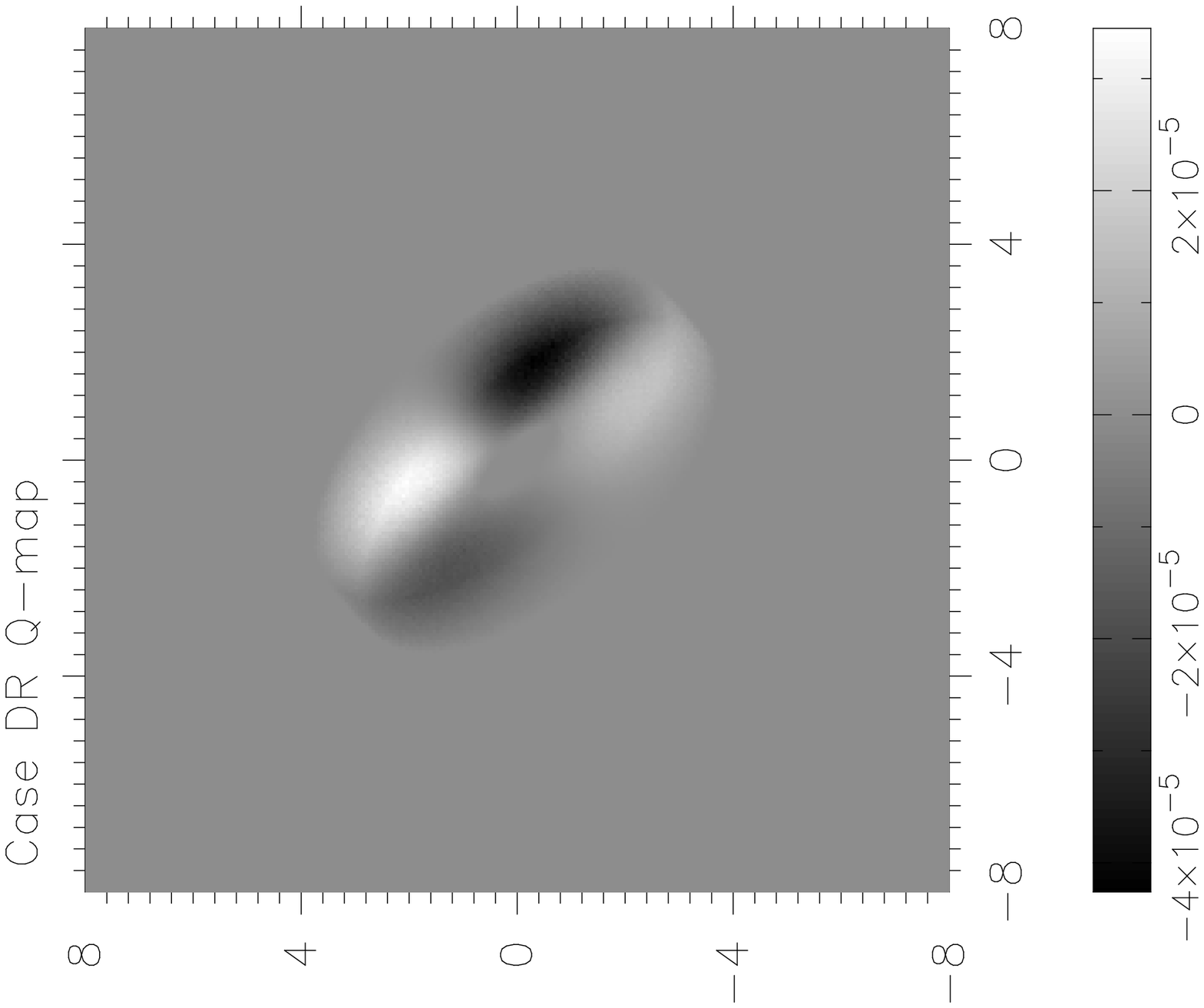}
 \end{minipage}
 \begin{minipage}{0.48\linewidth}
   \includegraphics[angle=-90,width=\linewidth]{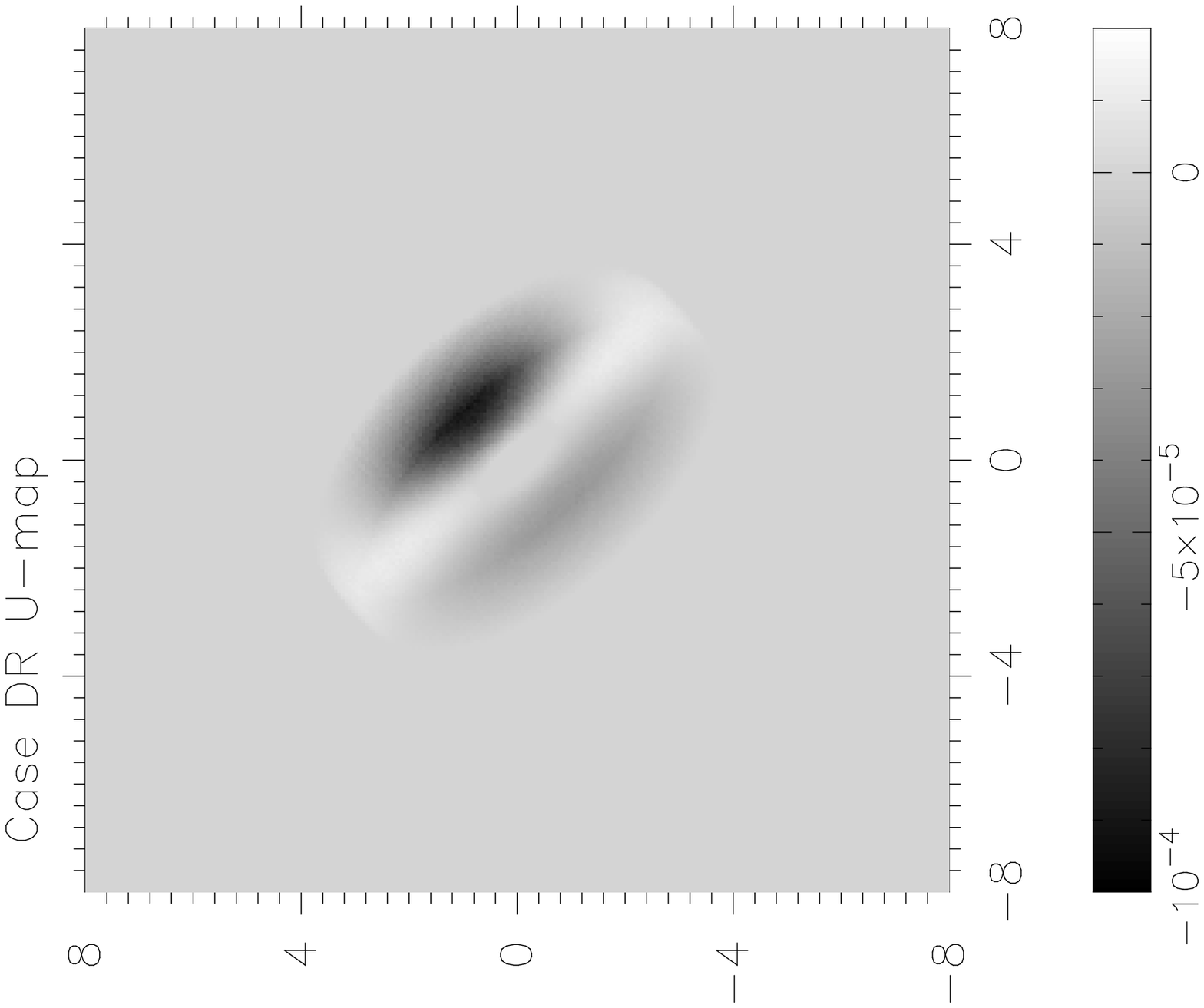}
 \end{minipage}
\end{center}
\caption{ \label{resultQU}
The comparison between the $Q$, $U$ maps observed in optical bands (Michel et al. 1991) 
and those of Case DR ($b=0.6$, $\beta=0.2$). 
The upper row is an observation, and the lower is the model.
The $Q$ map (left) is N/S polarized light minus E/W polarized light, 
the $U$ map (right) is NW/SE polarized light 
minus NE/SW polarized light.
}
\end{figure}

We reproduced the maps of Stokes parameters, 
Q-map and U-map in Fig.~\ref{resultQU}, 
which are compared with the observation (Michel et al. 1991). 
Because interaction of the filaments and the nebula flow changes 
the direction of the magnetic field, 
comparison may be limited within the area of $1^\prime$ around the pulsar. 
The shape of hourglass (Michel et al. 1991) is well reproduced in our Q map. 
It follows from this fact that the average magnetic field is toroidal. 
Because depolarization occurs along the major axis on which U has positive value, 
the absolute value of U is small (see Fig.~\ref{resultQU}). 
Therefore, U-map is easy to be influenced by disturbances, 
and therefore comparison of the observations to the model is difficult to make. 
 
\section{Summary and Discussion}
\label{discus}
We calculate the synchrotron polarization properties from a relativistically expanding disk 
which has both the toroidal mean field component and the disordered component. 
We compared our results with observations of the Crab Nebula. 
The distribution function of plasma particles 
and the total magnetic energy density 
$(B^2_0+B^2_1) / 8 \pi$ in the disk are given as a function of the distance $R$, 
according the KC flow with $\sigma = 0.003$. 
The best fit model reproducing the observed polarization 
indicates that the disordered magnetic field contributes $60 \%$ 
of the total magnetic energy.
This result is also supported by the fact that 
if the magnetic field is pure toroidal, 
then the nebula image becomes lip-shaped, 
while if there is such a disordered magnetic field, 
it becomes ring-shaped as observed. 
We also attempt to reproduce the front-back contrast of the ring by changing the flow velocity. 
The estimate of disordered field is not changed even if a high speed flow of $0.2c$ is introduced. 
It is shown that relativistic motion of the expanding nebula can be detected as asymmetry in P.A. and in distribution of depolarized regions. 

Nothing is known about wavelengths of the disordered magnetic field according to our study.
It is pointed out that series of magnetic neutral sheets 
in the wind can be collapsed at the shock (Lyubarsky 2003). 
If it is so, the disordered magnetic field may be formed 
in the process of magnetic reconnection. 
The scale lengths of the disordered field is thought of 
as $c/\Omega \approx 1.6 \times 10^8$ or less. 

Recently, relativistic magneto-hydrodynamic (RMHD) simulations
are done for the flow after the termination shock of the pulsar wind, 
and formation mechanism of the disk and the jets is discussed 
(Komissarov \& Lyubarsky 2003, Del Zanna et al. 2004).
Because of the axisymmetry, the magnetic field is assumed to be pure toroidal in those simulations.
Therefore, even though vortex is formed in the meridional plane, 
it does not cause significant depolarization (Bucciantini et al. 2005, Del Zanna et al. 2006). 

It is notable that if the disordered field dominates, 
then the magnetic field contributes more as pressure than tension in the flow dynamics, 
as compared with the case of pure toroidal field, 
which is assumed in the previous RMHD simulations. 
As a result, magnetic pinch effect may be overestimated 
in the previous simulations. 
If the disordered magnetic field is taken into account, 
the jets may be weaken, 
and magnetic pressure may cause acceleration of the disk flow.

Hickson (1990) finds highest polarization of $\sim 60 \%$ 
in a region $1^\prime.8$ apart from the pulsar in a south-south-east direction 
and the eastern bay. 
For these regions, very ordered magnetic field is expected. 
However, polarization degrees of $35 \%$ for the wisp1, $23 \%$ for the wisp3 (Hickson 1990), 
and $\sim 30 \%$ of the wisp3 (Scargle 1971) seem to be consistent with $b \sim 0.6$. 

\begin{figure}
\begin{center}
\includegraphics[clip,width=8cm]{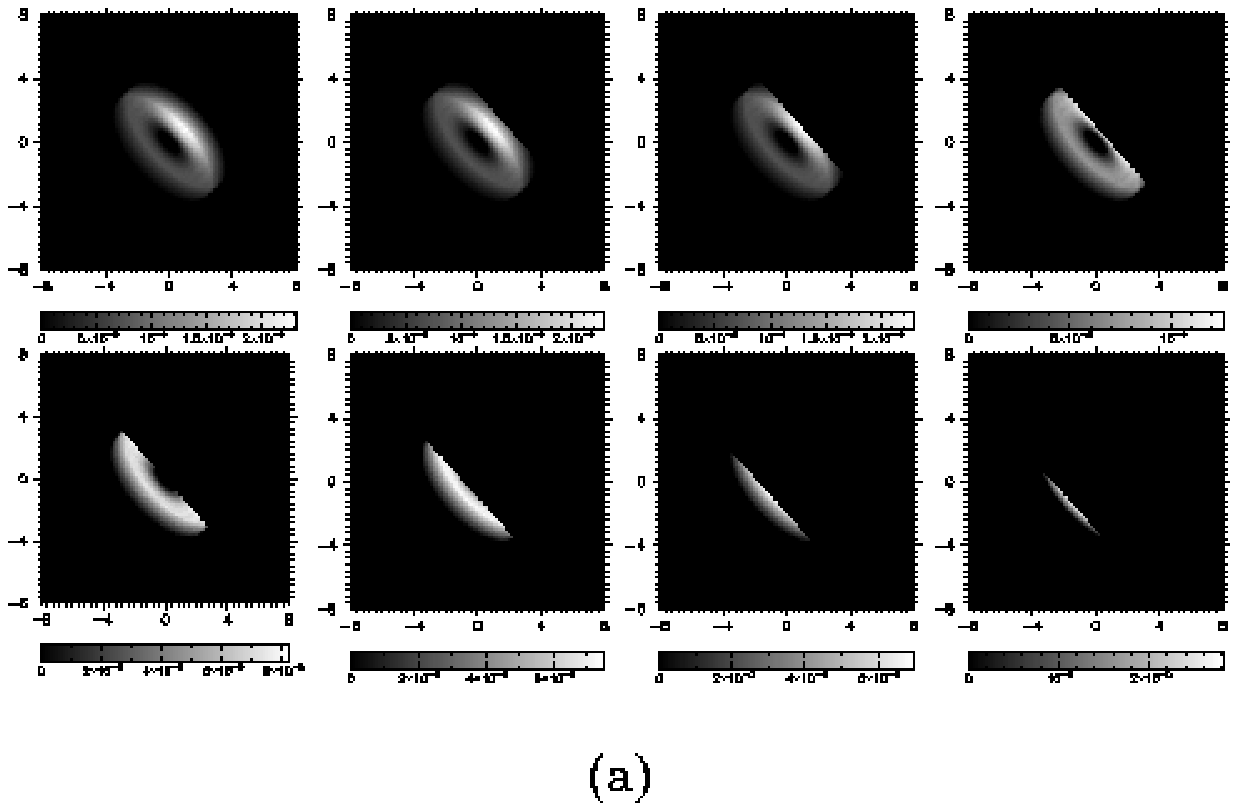}\\
\includegraphics[clip,width=4cm]{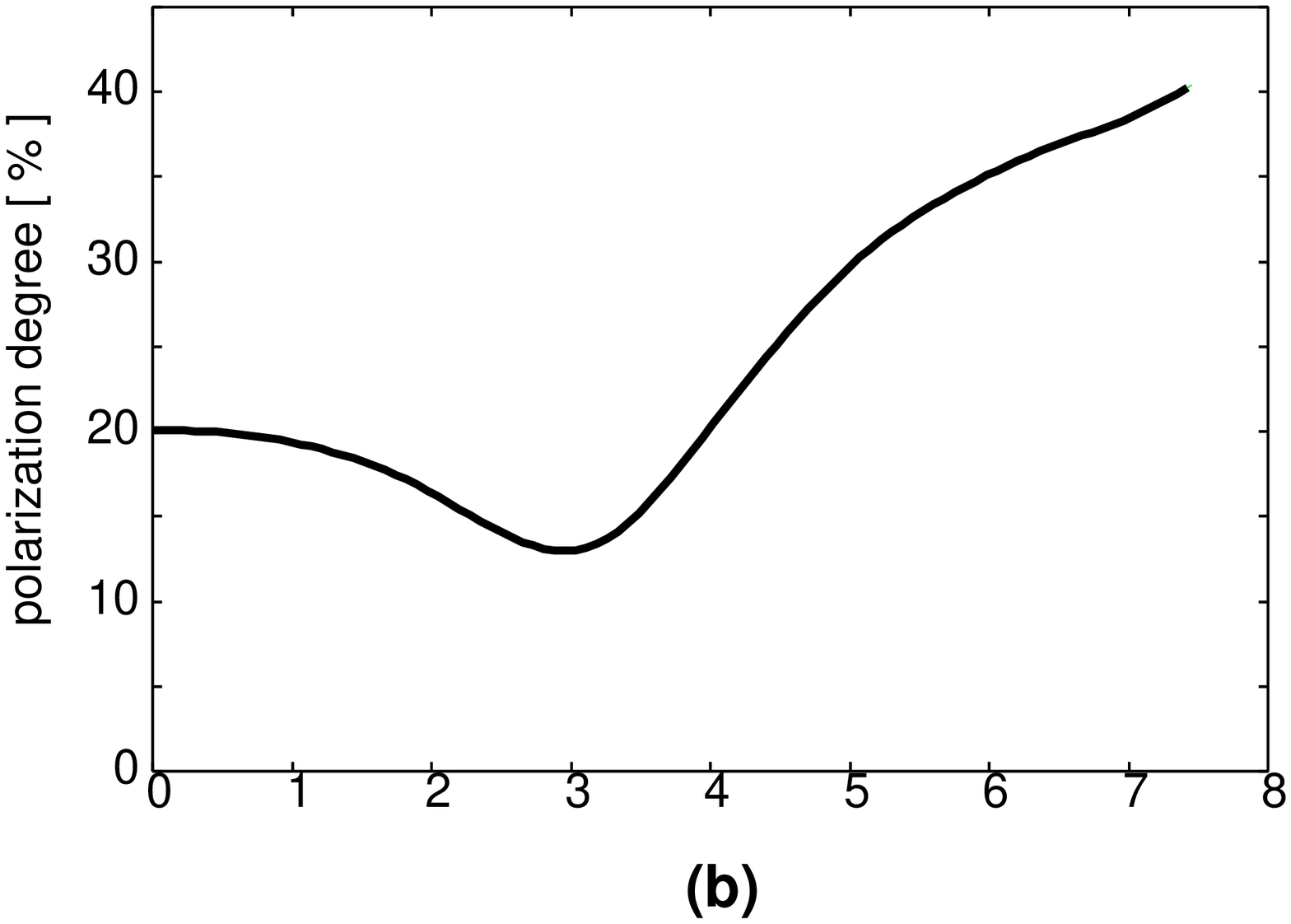}
\includegraphics[clip,width=4cm]{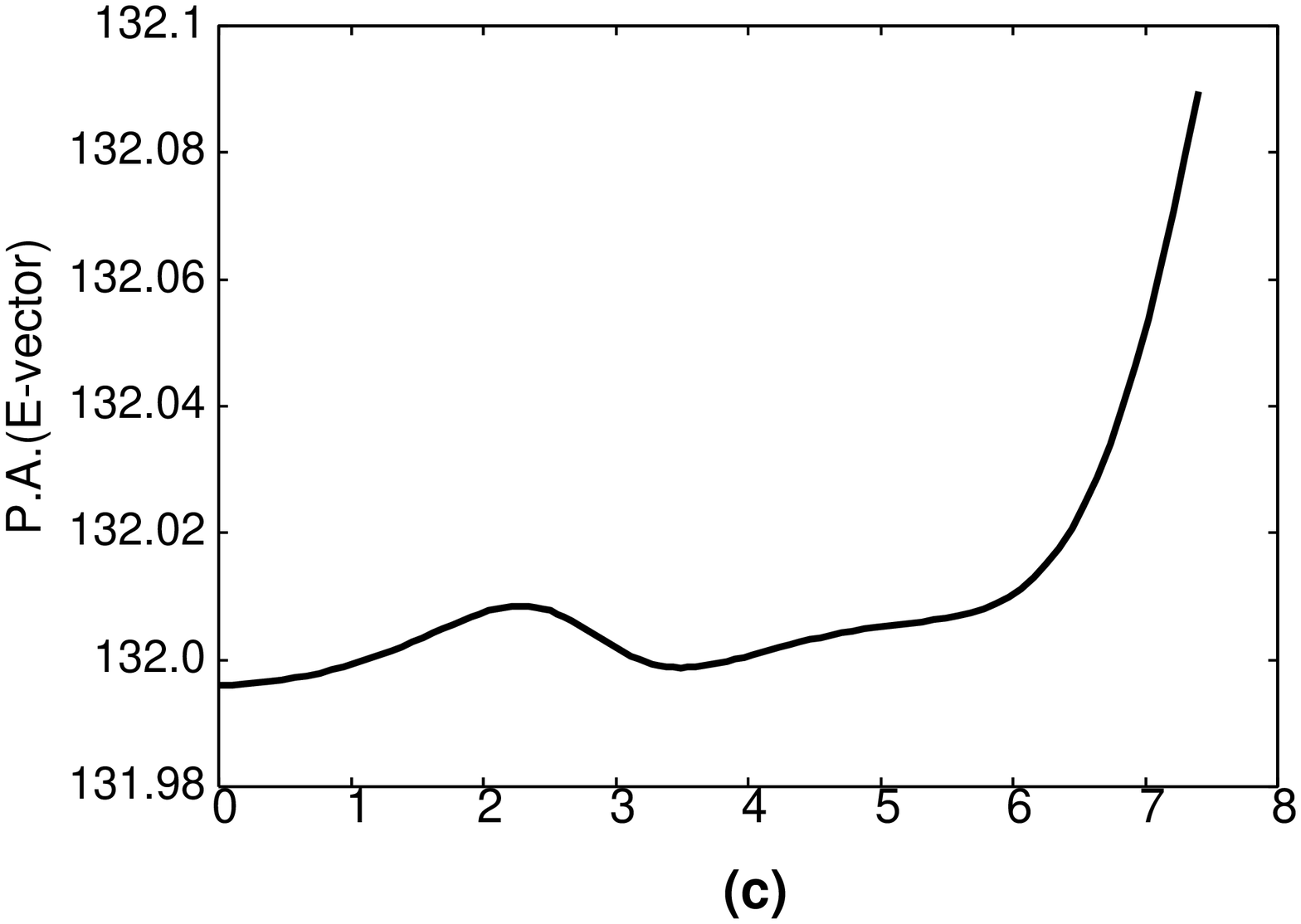}
\end{center}
\caption{ \label{enpei1}
Polarization at a lunar occultation of the Crab Nebula (disappearance phase). 
We use Case DR, $b=0.6$, $\beta=0.2$. 
(a)Appearance of luna occultation from the northwest to the southeast along symmetrical axis of the nebula. 
(b)Change of averaged polarization degree. 
(c)Change of P.A. 
The labels 0$\sim$7 of the horizontal axis of panel (b) and (c) 
correspond to eight figures from on the upper left to lower right of sub-panels in (a). 
}
\end{figure}

\begin{figure}
\begin{center}
\includegraphics[clip,width=8cm]{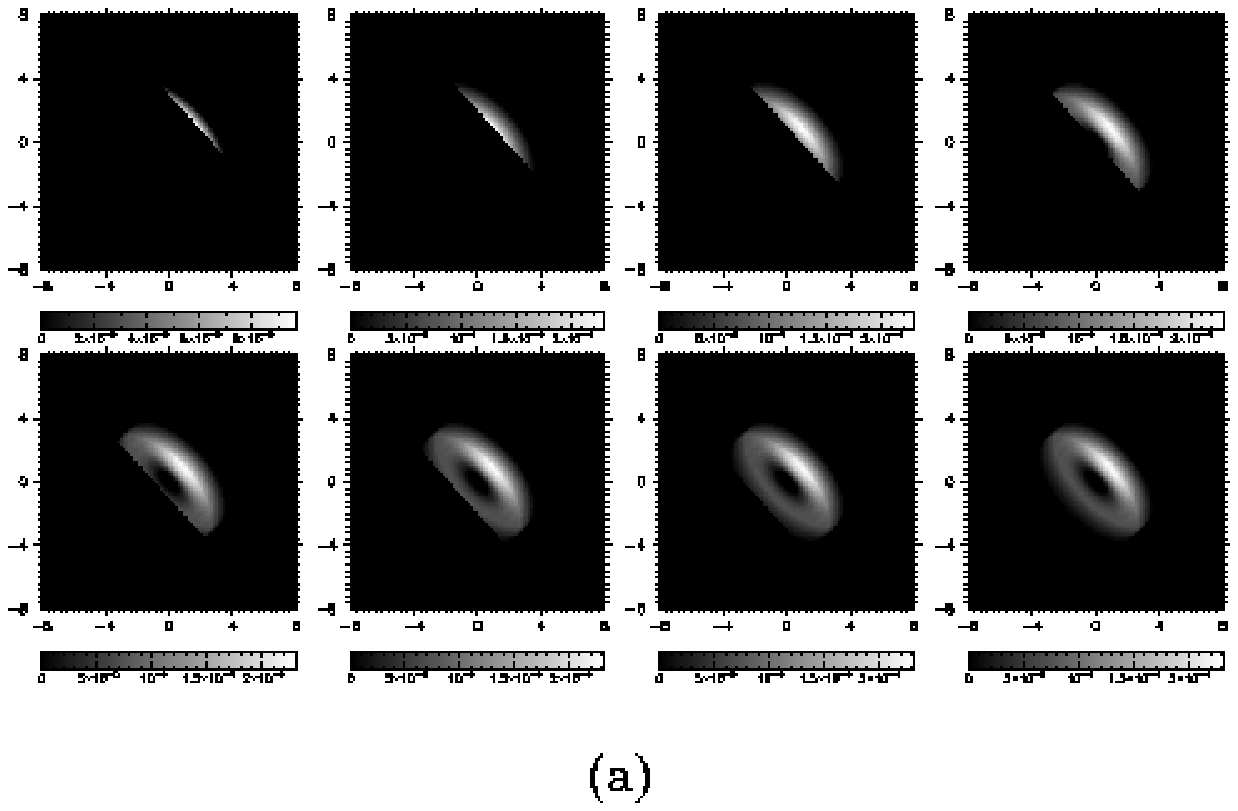}\\
\includegraphics[clip,width=4cm]{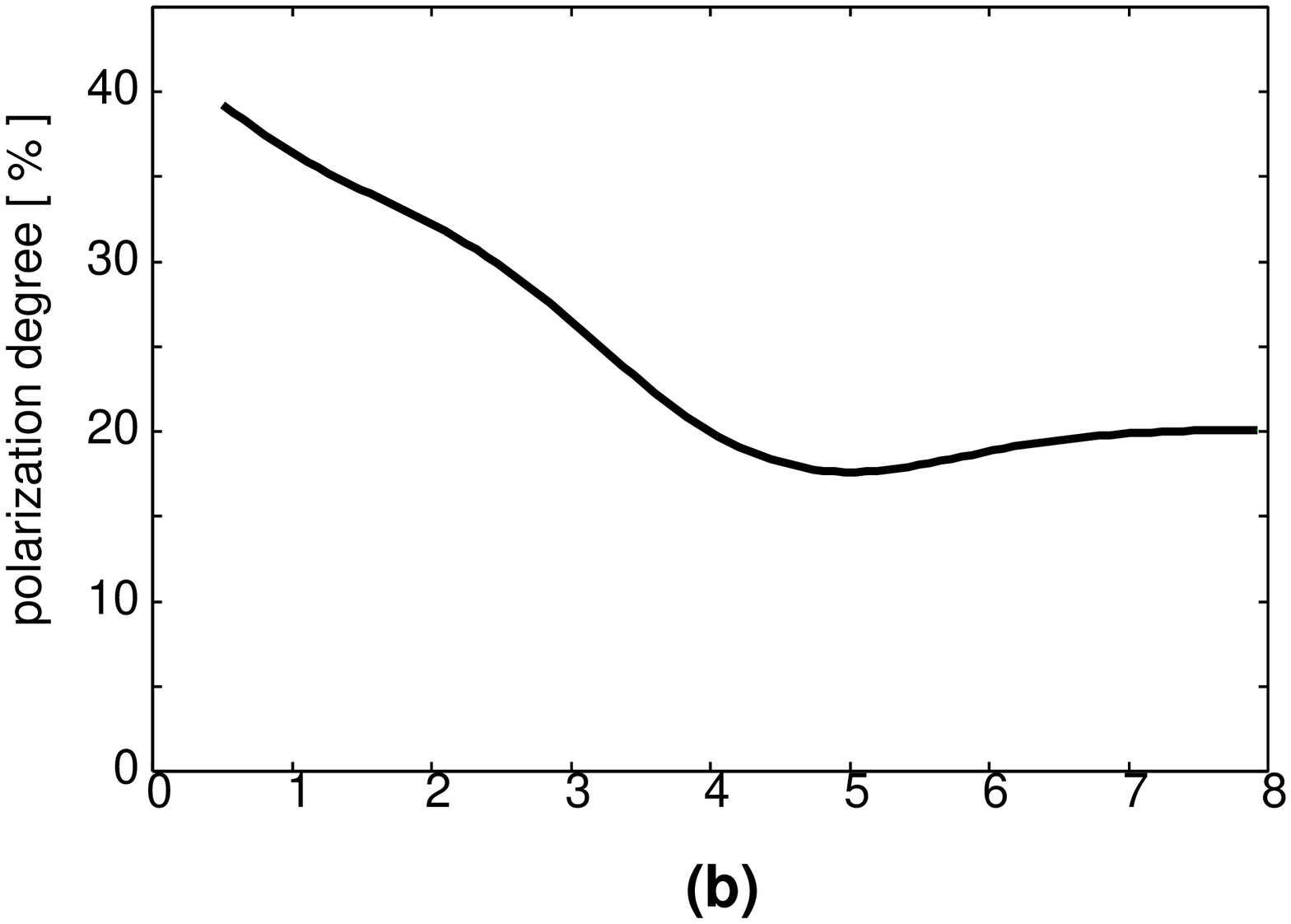}
\includegraphics[clip,width=4cm]{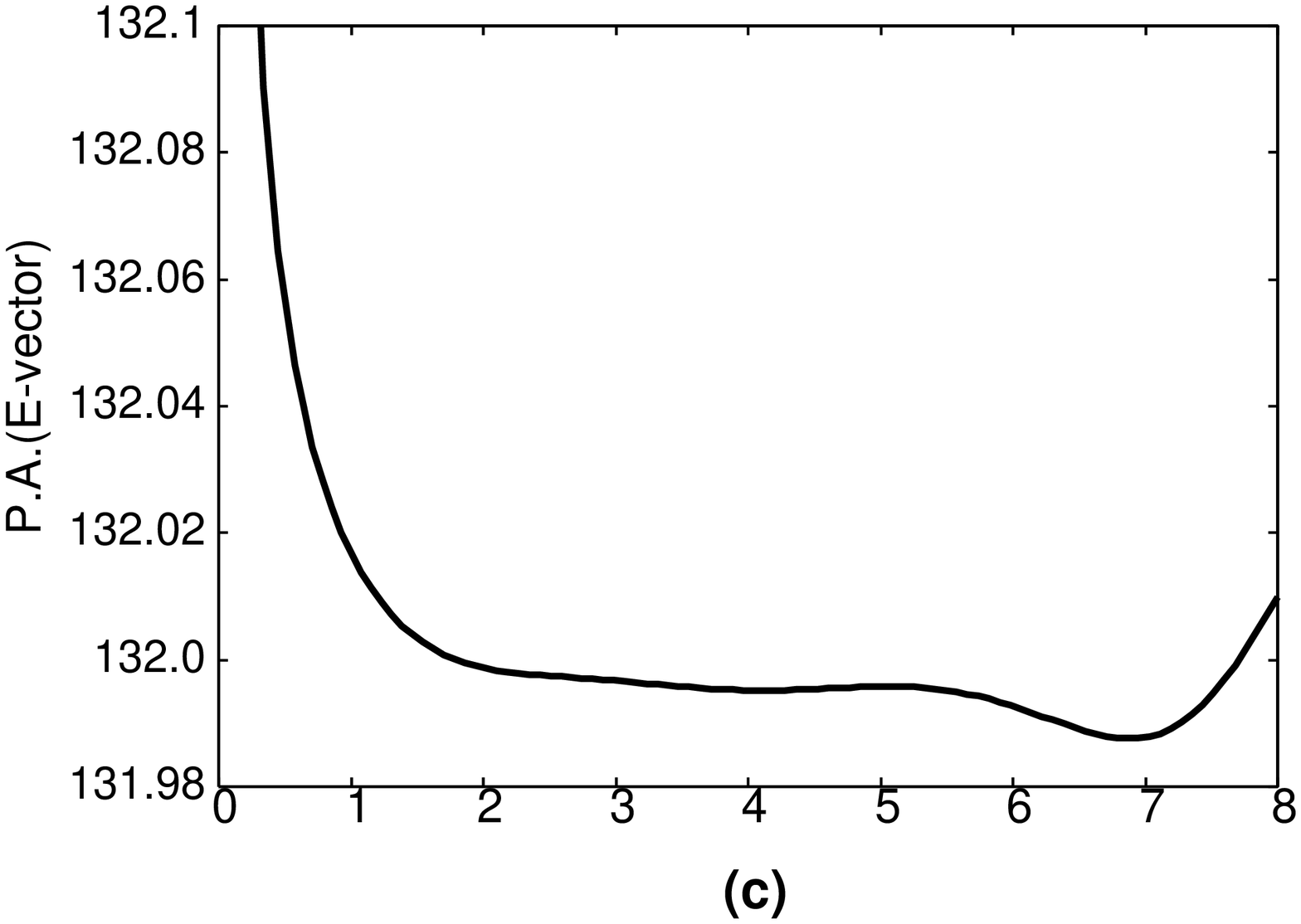}
\end{center}
\caption{ \label{enpei2}
The same as Fig.~\ref{enpei1}, but for the appearance phase. 
The labels 1$\sim$8 of the horizontal axis of panel (b) and (c) 
correspond to eight figures from on the upper left to lower right of sub-panels in (a). 
}
\end{figure}

For a future possibility, 
we simulated change of polarization for the entire Crab Nebula 
through a lunar occultation. 
Fig.~\ref{enpei1} shows the results when the nebula is 
gradually hidden, 
while Fig.~\ref{enpei2} shows a recovery phase. 
The direction of the occultation can be seen in the images of the figures.  
In the disappearance phase (Fig.~\ref{enpei1}), 
the total polarization degree decreases at first 
because of hiding highly polarized parts, 
and then it is minimized and increases. 
The late phase will not be observed 
because the south-east part of the ring is dim, 
and the radiation from the jet will dominate.  
In the reappearance phase, 
the polarization degree start with a very high state 
$\sim 40 \%$ and gradually recovers. 
There will be no significant change of position angle. 

\section*{Acknowledgments}
We would like to thank Dr. S. Gunji, Dr. H. Sakurai, Dr. K. Mori, and Dr. K. Ioka for fruitful discussions.
This work was supported Grant-in-Aid for Scientific Research 
from the Ministry of Education, Culture, Sports, Science and Technology of Japan (15540227).

\label{lastpage}

\begin{thebibliography}{}
\bibitem[\protect\citeauthoryear{Bj\"{o}rnsson}{1982}]{Bj}
   Bj\"{o}rnsson, C.-I., 1982, ApJ, 260, 855
\bibitem[\protect\citeauthoryear{Blandford \& K\"{o}nigl}{1979}]{BK}
   Blandford, R.D., \& K\"{o}nigl, A., 1979, ApJ, 232, 34
\bibitem[\protect\citeauthoryear{Begelman \& Li}{1994}]{BLi}
   Begelman, M. C., \& Li, Z.-Y., 1994, ApJ, 426, 269
\bibitem[\protect\citeauthoryear{Bogovalov}{1998}]{Bog}
   Bogovalov S. V., 1998, Astronomy Letters, 24, 321-331
\bibitem[\protect\citeauthoryear{Bucciantini, Del Zanna, Amato \& Volpi}{2005}]{BDAV} 
   Bucciantini, N., Del Zanna, L., Amato, E., \& Volpi, D., 2005, A \& A, 443, 519B 
\bibitem[\protect\citeauthoryear{Coroniti}{1990}]{Cor}
   Coroniti, F.V., 1990, ApJ, 349, 538
\bibitem[\protect\citeauthoryear{Del Zanna, Amato \& Bucciantini}{2004}]{DAB}
   Del Zanna, L., Amato, E., \& Bucciantini, N., 2004, A \& A, 421, 1063
\bibitem[\protect\citeauthoryear{Del Zanna, Volpi, Amato \& Bucciantini}{2006}]{DVAB}
   Del Zanna, L., Volpi, D., Amato, E., \& Bucciantini, N., 2006, A \& A, 453, 621D
\bibitem[\protect\citeauthoryear{Hester, Mori, Burrows, Gallagher, Graham, Halverson, Kader, Michel \& Scowen}{2002}]{He}
   Hester, J. J., Mori, K., Burrows, D., Gallagher, J. S., Graham, J. R., Halverson, M., Kader, A., Michel, F. C., Scowen, P., 2002, ApJ, 577, 49
\bibitem[\protect\citeauthoryear{Hickson \& van der Bergh}{1990}]{HvdB}
   Hickson, P., \& van der Bergh, S., 1990, ApJ, 365, 224
\bibitem[\protect\citeauthoryear{Kennel \& Coroniti}{1984}]{KCa}
   Kennel, C.F., \& Coroniti, F.V., 1984a, ApJ, 283, 694
\bibitem[\protect\citeauthoryear{Kennel \& Coroniti}{1984}]{KCb}
   Kennel, C.F., \& Coroniti, F.V., 1984b, ApJ, 283, 710
\bibitem[\protect\citeauthoryear{Kirk \& Skjaeraasen}{2003}]{KiSk}
   Kirk J.G., \& Skjaeraasen O., 2003, ApJ, 591, 366
\bibitem[\protect\citeauthoryear{Komissarov \& Lyubarsky}{2003}]{KL1}
   Komissarov, S.S., \& Lyubarsky, Y.E., 2003, MNRAS, 344, L93
\bibitem[\protect\citeauthoryear{Korchakov \& Syrovat-skii}{1962}]{KS} 
   Korchakov, A.A., \& Syrovat-skii, S.I., 1962, SvA, 5, 678
\bibitem[\protect\citeauthoryear{Lyutikov, Pariev \& Blandford}{2003}]{LPB}
   Lyutikov, M., Pariev, V.I., \& Blandford, R.D., 2003, ApJ, 597, 998
\bibitem[\protect\citeauthoryear{Lyuvarsky}{2003}]{Lyu}
   Lyuvarsky, Y. E., 2003, MNRAS, 345, 153-160
\bibitem[\protect\citeauthoryear{Michel}{1969}]{Mic}
   Michel, F. C., 1969, ApJ, 158, 727
\bibitem[\protect\citeauthoryear{Michel, Scowen, Dufour \& Hester}{1991}]{MSDH}
   Michel, F.C., Scowen, P.A., Dufour, R.J., \& Hester, J.J., 1991, ApJ, 368, 463
\bibitem[\protect\citeauthoryear{Mori, Burrows, Hester, Pavlov, Shibata \& Tsunemi}{2004}]{Mo}
   Mori, K., Burrows, D. N., Hester, J. J., Pavlov, G. G., Shibata, S., \& Tsunemi, H., 2004, ApJ, 609, 186
\bibitem[\protect\citeauthoryear{Ng \& Romani}{2004}]{NgR}
   Ng, C. -Y., \& Romani, R. W., 2004, ApJ, 601, 479
\bibitem[\protect\citeauthoryear{Okamoto}{2002}]{Ok}
   Okamoto, I., 2002, ApJ, 573, L31
\bibitem[\protect\citeauthoryear{Oort \& Walraven}{1956}]{OW}
   Oort, J. H., \& Walraven, T., 1956, Bull. Astr. Inst. Netherlands, 12, 285
\bibitem[\protect\citeauthoryear{Rybicki \& Lightman}{1979}]{RL}
   Rybicki, G.B., \& Lightman, A.P., 1979, Radiative Processes in Astrophysics, Wiley, New York
\bibitem[\protect\citeauthoryear{Scargle}{1971}]{Sc}
   Scargle, J.D., 1971, Nature Phys. Sci., 230, 37
\bibitem[\protect\citeauthoryear{Schmidt, Angel \& Beaver}{1979}]{SAB}
   Schmidt, G.D., Angel, J.R.P., \& Beaver, E.A., 1979, ApJ, 227, 106
\bibitem[\protect\citeauthoryear{Shibata, Tomatsuri, Shimanuki, Saito \& Mori}{2003}]{STSSM}
   Shibata, S., Tomatsuri, H., Shimanuki, M., Saito, K., \& Mori, K., 2003, MNRAS, 346, 841
\bibitem[\protect\citeauthoryear{Tomimatsu}{1994}]{Tomi}
   Tomimatsu A., 1994, PASJ, 46 123
\bibitem[\protect\citeauthoryear{Velusamy}{1985}]{Ve}
   Velusamy, T., 1985, MNRAS, 212, 359
\bibitem[\protect\citeauthoryear{Vlahakis}{2004}]{Vl}
   Vlahakis, N., 2004, ApJ, 600, 324
\bibitem[\protect\citeauthoryear{Weisskopf, Silver, Kestenbaum, Long \& Novick}{1978}]{We1}
   Weisskopf, M. C., Kestenbaum, H. L., Long, K. S., Novick, R., Silver, E. H., 1978, ApJ, 220, 117
\bibitem[\protect\citeauthoryear{Weisskopf, Hester, Tennant, Elsner, Schulz, Marshall, Karovska, Nichols, Swartz, Kolodziejczak \& O'Dell}{2000}]{We2}
   Weisskopf, M. C., Hester, J. J., Tennant, A. F., Elsner, R. F., Schulz, N. S., Marshall, H. L., Karovska, M., Nichols, J. S., Swartz, D. A., Kolodziejczak, J. J., \& O'Dell, S. L., 2000, ApJ, 536, 81
\bibitem[\protect\citeauthoryear{Wilson}{1972}]{Wil}
   Wilson, A. S., 1972, MNRAS, 157, 229
\end{thebibliography}
\end{document}